%% file: main.tex
\documentclass[acmtog,nonacm, table, supertabular, hyperref,dvipsnames]{acmart}

\usepackage{stfloats} % Put in preamble to enable [b] for starred floats
\usepackage{placeins}
\usepackage{xcolor}
\usepackage[ruled,vlined]{algorithm2e}
\usepackage{booktabs}
\usepackage{multirow}
\usepackage{wrapfig}

\usepackage{utfsym}
\usepackage{longtable}

\usepackage[table]{xcolor} % For rowcolors / lightcornflower

\AtBeginDocument{%
  }

\setcopyright{none}
\acmVolume{0}
\acmNumber{0}
\acmArticle{0}
\acmMonth{0}

\input{sections/pre.tex}

\begin{document}

\title{\emoji{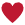} Handles: Decimation for Deformation Handles with Compact Support and Low Memory Footprints} 

\author{David IW Levin}
\affiliation{%
  \institution{University of Toronto and NVIDIA}
  \country{Canada}
}

\author{Paul Kry}
\affiliation{%
  \institution{McGill University}
  \country{Canada}
}

\author{Kartic Subr}
\affiliation{%
 \institution{University of Edinburgh}
 \country{UK}
}

\author{Ryan Schmidt}
\affiliation{
    \institution{gradientspace}
    \country{Canada}
}
\author{Etienne Vouga}
\affiliation{%
  \institution{University of Texas, Austin}
  \country{USA}
}

\author{Teseo Schneider}
\affiliation{%
  \institution{University of Victoria}
  \country{Canada}
}

\renewcommand{\shortauthors}{Levin et al.}

\begin{abstract}
Estimating the deformation of solids via physical simulation is an important problem spanning fields such as computer animation, engineering and robotics. Such simulations are computationally expensive and scale poorly when the representation of an object is refined by increasing the level of discretization.
\emph{Reduced Order Methods (ROM)} offer computational savings by decreasing the number of degrees of freedom, for example by using \emph{handles} that control groups of vertices. We present the first decimation-based algorithm for computing a sparse, compactly supported set of deformation handles. 
The crux of our method utilizes iterative algebraic simplification to optimize handle deformation to match any input deformation, such as linear vibration modes. This applies to any volumetric input mesh, including those with high genus or porous features, since we do not alter the geometry. We also devise an efficient algorithm to compute and update compact supports and their associated weights. We  leverage compact support to develop an efficient, reduced-cubature computation scheme. Once optimized, our handles offer a memory-efficient solution while enabling real-time elastodynamics simulation of complex geometry. We show real-time performance on a variety of tetrahedral meshes with up to $796,623$ tetrahedra.
\end{abstract}

\begin{teaserfigure}
  \centering
  \includegraphics[width=\textwidth]{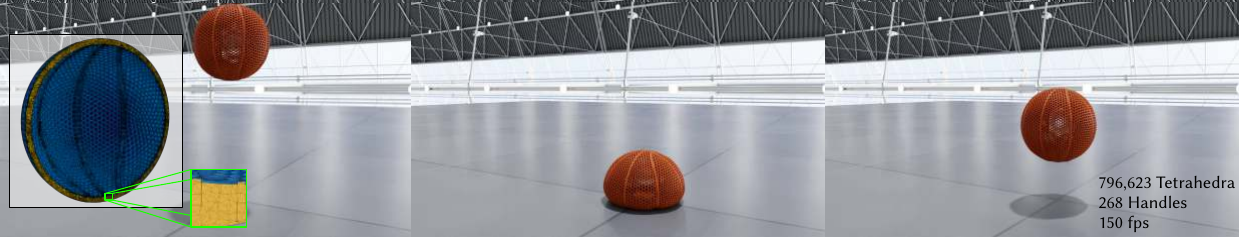}
  \caption{Our handle decimation algorithm reduces the degrees-of-freedom of this metamaterial basketball from 200,000 vertices to $268$ handles, maintaining $5\%$ error for the first $25$ linear modes. Coupled with our reduced cubature scheme this enables \emph{nonlinear} elastodynamics (ARAP, Young's Modulus $1e^7$ Pa) simulation at $150$ steps per second.}
  \label{fig:teaser}
\end{teaserfigure}

\maketitle

\section{Introduction}
\input{sections/introduction}

\section{Related Work}
\input{sections/related.tex}

\section{Preliminaries}
\input{sections/preliminaries}

\section{Methods}
\input{sections/methods}

\section{Results}
\input{sections/results}

\section{Conclusions}
\input{sections/conclusion.tex}

\bibliographystyle{ACM-Reference-Format}
\bibliography{lovehandles}

\appendix
\renewcommand{\thesection}{A.\arabic{section}}
\input{sections/appendices}

%\begin{acks}
%The Bellairs Workshop on Computer Animation was instrumental in the %conception of the research presented in this paper.
%\end{acks}

\end{document}

%% file: sections/pre.tex
\newcommand{\nummodes}{k}
\newcommand{\hsupp}{H}
\newcommand{\vertices}{\mathcal{V}}  % PGK made mathcal to make it a set
\newcommand{\vertex}{v}

\newcommand{\tets}{\mathcal{T}} % PGK made math cal to make it a set
\newcommand{\numverts}{|\vertices|}

\newcommand{\hori}{\mathbf{C}}
\newcommand{\fieldv}[1]{\mathbf{u}^{#1}}

\newcommand{\Taffine}{\mathbf{T}}

\newcommand{\supp}[1]{\mathcal{S}^{-1}(#1)}
\newcommand{\invsupp}[1]{\mathcal{S}(#1)}
\newcommand{\newsupp}[1]{\mathcal{S}^{*-1}(#1)} 
\newcommand{\wv}{\mathbf{W}}

\newcommand{\wfun}{w}
\newcommand{\errv}[1]{e_{#1}}
\newcommand{\errg}{\mathcal{E}}
\newcommand{\errgth}{\varepsilon}
\newcommand{\errdelta}[1]{\delta_{#1}}
\newcommand{\mscale}[1]{\sigma_{#1}}
\newcommand{\pqueue}{\mathcal{Q}}

\newcommand{\handle}{h}
\newcommand{\sphandle}{h} % PGK: sp = "section preliminaries" handle index, was i
\newcommand{\disphandle}{h} % PGK: displacements definition handle index, where handles are centroids, was c
\newcommand{\handleset}{\mathcal{H}}
\newcommand{\nhandles}{|\handleset|}

\newcommand{\volgraph}{\mathcal{G}}

\newcommand{\refx}{\mathbf{X}}
\newcommand{\defx}{\mathbf{x}}

\newcommand{\grad}{\nabla}

\usepackage{xspace}
\newcommand{\POU}{PoU\xspace}
\newcommand{\selectW}{P_h}
\newcommand{\selectV}{S_h}
\newcommand{\Uaffine}{\mathbf{U}}

\makeatletter
\renewcommand{\sectionautorefname}{\S\@gobble}
\makeatother
\makeatletter
\renewcommand{\subsectionautorefname}{\S\@gobble}
\makeatother
\makeatletter
\renewcommand{\subsubsectionautorefname}{\S\@gobble}
\makeatother
\makeatletter
\renewcommand{\appendixautorefname}{\@gobble}
\makeatother

\definecolor{teseoCol}{rgb}{.15, .68, .38}
\definecolor{paulCol}{rgb}{.99, .00, .99}
\definecolor{karCol}{rgb}{.39, .10, .99}
\definecolor{daveCol}{rgb}{.99, .0, .0}

\colorlet{lightcornflower}{CornflowerBlue!15}

\newcommand{\emoji}[1]{\protect\includegraphics[height=0.8em]{#1}}

%% file: sections/introduction.tex
\begin{figure*}
\begin{center}
    \includegraphics[width=\linewidth]{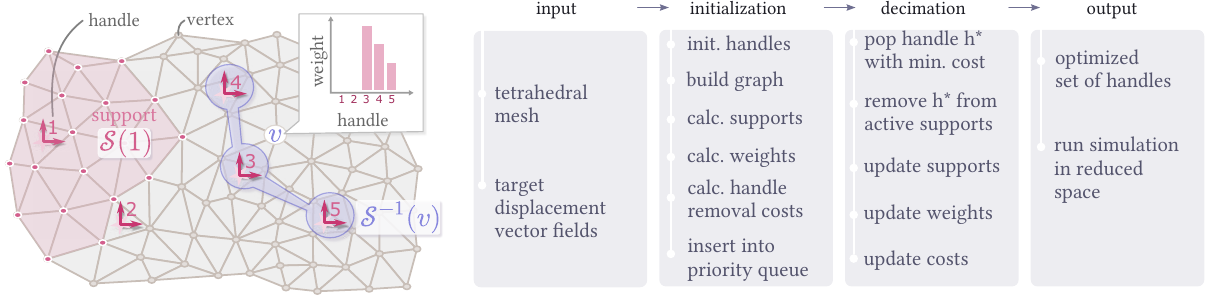}
    \caption{\label{fig:overview} We reduce the degrees of freedom of a tetrahedral mesh to a set of handles (pink frames number $1-5$). The effect of each handle on vertex  $v$ of the mesh is specified via a weight function. The handles have compact support ($\invsupp{1}$ is shaded in pink). An overview of our algorithm is shown on the right.}
    \Description{Diagram showing supports, and data flow in algorithm.}
\end{center}    
\end{figure*}

Simulating the physics of deforming solids plays a key role across applications such as visual effects, interactive games and robotics. Often, computational efficiency is a crucial consideration. For example,  interactive applications place strong constraints on the time taken to simulate each frame, since its lower bound is the standard real-time rendering frame rate of 30 frames-per-second. In robotics applications that use reinforcement learning, large volumes of simulation data are required, hundreds of simulation steps are expected per second. Typical robotics simulation frameworks include only rudimentary non-rigid simulations. Even in the absence  of such application-induced constraints, efficient simulation enables  simulation at high rates, or small time-steps, with many benefits such as reduced damping, more robust  collision handling, etc.

Accelerating the simulation of elastic solids is a well-researched problem across mathematics, engineering, and computer science. Amongst the many approaches that have been devised, reduced-order models (ROMs) have shown great promise due to their effectiveness for fast simulation, their applicability in conjunction with newer representations (neural methods) and implementation on GPUs. Specifically, \emph{hyper-reduced methods} lead to large savings since they reduce the kinematic degrees-of-freedom (DOFs) and the complexity of cubature. 
%ROMs have their origins in mechanical engineering in the 1960s\KS{citation?} but continue to remain relevant in conjunction with newer representations, such as neural methods.
%
Handle-based ROMs represent the kinematics of an object using the linear-blend skinning formalization first proposed in computer graphics for character animation~\cite{Badler1982Modelling}. These methods have an uncanny ability to reproduce  complex motion with relatively few degrees-of-freedom. Unlike volumetric meshes, handles effectively represent a Cosserat-like medial motion, granting equal efficacy for both thin and volumetric solids, meaning they require only the specification of handle number, handle position, and associated weight functions across an object to produce compelling simulations of a wide range of complex geometry.

Automatically determining optimal handles and weights for arbitrary geometry is a challenging and open problem. Previous methods to produce simulation-ready handle-based discretizations either (1) assume that handle positions are known and compute the weights post hoc, or (2) use clustering to first find handle positions and then compute the weights. Such approaches suffer from poor memory efficiency~(Table~\ref{tab:brandt_comparison}) for complex simulations with dynamic boundary conditions. The calculation of weights based on eigenanalysis assumes that all handles are positioned at the same point in an object's reference space. This typically leads to dense and global weights increasing coupling across all parts of the object. The introduction of new boundary conditions can therefore lock the system, causing unnaturally stiff deformations. Dense, global weights require storage on the order of the number of vertices in the input mesh, and so only a few such weights can be effectively stored, limiting the expressivity of the final ROM.

Sparse regularizers~\citep{brandt2017compressed} address some of these challenges but do not guarantee compact support, meaning issues with coupling distant parts of an object still occur. Cluster-first approaches~\citep{James2005} cluster on metrics that do not necessarily produce good handle placement. For instance, works that cluster based on compliance (inverse of Young's modulus) are at risk of avoiding placing handles in stiff, but thin, parts of an object that should deform. More sophisticated clustering methods based on modal features may connect distant parts of the mesh that move in similar ways but are far apart on the input shape.

% PKG: Geko and Cable figure moved to section 2 where they are referenced in the body.
% but I note hte discussion of Cosserat-like stuff in the paragraph above could call for this figure to be moved back (with an appropriate Fig reference in this section.
We solve this problem taking inspiration from methods in iterative mesh decimation which collapse one mesh primitive (for triangle meshes, usually an edge) at a time. The edge is chosen to induce the least error in a geometric sense, relative to the original mesh. We adopt this hill-descending approach to handle placement and weight computation~(\autoref{fig:overview}). We begin with an overprescribed set of handles and weights, and at each iteration of our method, we remove one handle and recompute the weights for the others affected. Our key observation is that this recomputation is possible at practical rates and scales. Crucially, our approach enables the construction of weight functions that are shape- and deformation-aware, have guaranteed compact support (i.e., weights that are both sparse and limited to a contiguous region near their associated handle) and a low per-vertex memory storage. We then exploit compact support to perform efficient decimation. 

Given a tetrahedral mesh and a set of target displacement (vector) fields, we propose a method that simultaneously searches for handle placement and associated handle weights towards minimizing an error associated with reconstructing the set of fields in the reduced space defined by the handles.
We then exploit the compact nature of our weight functions to compute reduced cubature rules and enable fast GPU-based simulation at speeds over 100 frames-per-second for most geometries.
We achieve this by shifting the computational burden to a pre-process that involves the iterative decimation algorithm. However, we show how this can yield interactive runtimes (simulation), across a wide range of geometries, which are well suited for the massive sample requirements of applications such as reinforcement learning and game engines~(Fig. \ref{fig:teaser}, Fig. \ref{fig:interactive}).

In summary, we make the following contributions:
\begin{itemize}
\item the first decimation-based algorithm for error bounded handle position and weight computation;
\item the first algorithm which produces compactly-supported weights with fixed, per-vertex memory storage;
\item and a spectrally informed method to compute reduced cubature rules for our compactly supported handle discretization.
\end{itemize}
We also demonstrate the utility of the handle-based output on a wide variety of simulations using a handle-based elasticity simulator implemented on the GPU.

%% file: sections/related.tex
Reduced-order models (ROMs), introduced by \citet{PentlandWilliams1989}, have become a workhorse tool in computer simulation with a correspondingly rich body of related work. Over the years, modal simulation has been extended to nonlinear constitutive models~\citep{ChoiKo2005,Barbic2005}, fracture~\citep{sellan2023breaking}, and handle-based discretizations~\citep{Gilles2011,Faure2011,benchekroun2023fast,trusty2023subspace,10.1145/3450626.3459753}, and has been sparsified via regularization~\citep{brandt2017compressed}. More recently, neural representations have shown promise for producing ROMs for generalized geometry representations~\citep{chen2023crom,chang2023licrom,Simplicits2024,xiang2026freeform,liao2026boundaryaware}.

\begin{figure}
\includegraphics[width=\columnwidth]{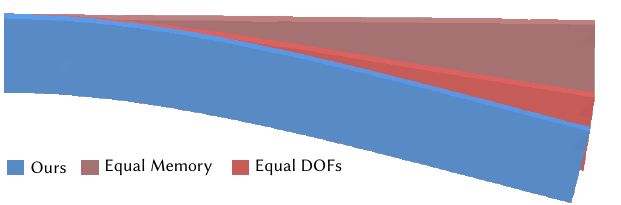}
\caption{Simulating a cantilevered beam with our method and linear modes, both computed without the included fixed boundary condition. Linear modes experience severe locking at equal memory usage and moderate locking at equal DOF count ($10 \times$ more memory usage) relative to ours.}
    \Description{Bending cantilevers pinned on one side.}

\label{fig:compare_linear}
\end{figure}

\begin{figure}
    \includegraphics[width=\columnwidth]{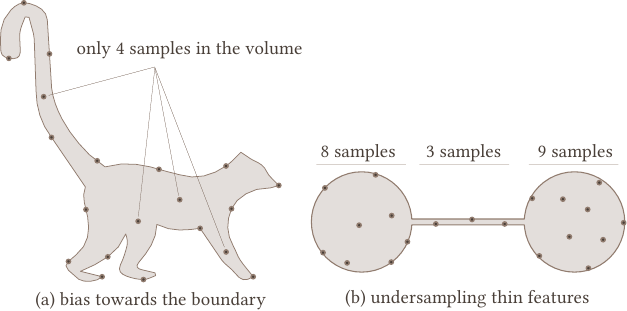}
    \caption{Farthest-point sampling, commonly used to place handles in objects, simultaneously undersamples both thin features and large volumetric regions depending on input geometry.}
    \Description{Diagram showing samples on a 2D monkey and a 2D barbell.}
    \label{fig:fps_bias}
\end{figure}

A remaining limitation of most reduced-order methods is that they produce globally supported basis or weight functions. Global support leads to locking under new boundary conditions~(\autoref{fig:compare_linear}) and leads to ``spooky action at a distance''~\citep{benchekroun2025force}. Other works~\citep{xian2019scalable,trusty2025sparse} produce compactly supported basis functions to avoid these issues; however, these are used with hierarchical full-space solvers rather than as standalone ROMs. \citet{10.1145/3197517.3201387} and \citet{li2019multi} create compactly-supported basis functions but do so using purely distance based farthest-point sampling which is not material-aware, suffers from density bias~(\autoref{fig:fps_bias})\ and, for the former, requires tuning a distance threshold to obtain a well-posed reduced space, all things our method avoids by construction since it is error-driven. 

Our approach is motivated by two seemingly disparate related works. The first is \citet{James2005}, which uses a clustering step, followed by a weight fitting step to compress vertex-based animations. The second is the mesh simplification/decimation work of \citet{Garland1997}. We observe that ROM constructions based on modal analysis~\citep{PentlandWilliams1989} or optimization~\citep{benchekroun2023fast} make compact support (a combination of locality and sparsity) difficult to encode. \citet{chen2019material} produces bases that decay exponentially but are not necessarily compactly supported, requiring both memory storage on the order of the mesh size for each basis function and potentially difficult-to-choose tolerances (or more complicated algorithms) to achieve fast linear algebra performance when the number of bases grows large. \citet{benchekroun2025force} achieves locality via force probability distribution which are hard to acquire and often hand-crafted. An alternative to solving a global optimization problem for computing reduced handle weights is to first compute handle positions via clustering, then, given those handle locations, compute optimal weights. For instance, \citet{James2005} uses mean-shift clustering to produce near-rigid clusters and then computes weights that best reconstruct a given animation. However, the computed weight functions are neither sparse nor compactly supported, suffering from all the issues discussed above. Later works~\citep{Faure2011,trusty2025sparse} produce compactly supported weights by adding additional constraints based on cluster boundaries. However, these methods cluster based on the compliance (inverse Young's modulus) of the input geometry, meaning that they undersample stiff but thin geometries (like a tree branch) which should deform. In general, designing an optimal clustering metric that respects compact support and deformation magnitude is an unsolved problem. 

Like~\citet{James2005}, our method also attempts to compute a number of handles, their positions in space, and a set of weights that reconstruct a set of $\nummodes$ target displacement fields. As \citet{chen2017dynamics} argue, a coarse discretization should replicate the low-energy modes of its high-resolution counterpart; therefore, we choose $\nummodes$ linear modes computed on a high-resolution input mesh as our targets. In this setting, clustering based on rigidity no longer holds, as the collection of modes is unlikely to be rigid at any point. Clustering based on modal features will group far-away mesh areas that oscillate in similar ways.

\begin{figure}[t]
\includegraphics[width=0.9\columnwidth]{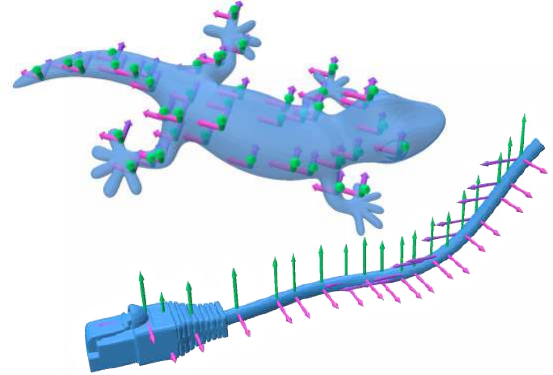}
\caption{Handle distributions visualized for Gecko (Top) and Ethernet Cable (Bottom). Our method concentrates handles in areas of high deformation (tail, legs of Gecko) and correctly positions handles along the medial axis of  the thin cable geometry.}
\Description{Frames visualized within transparent model visualizations.}
\label{fig:dist}
\end{figure}

\begin{figure}
\includegraphics[width=\columnwidth]{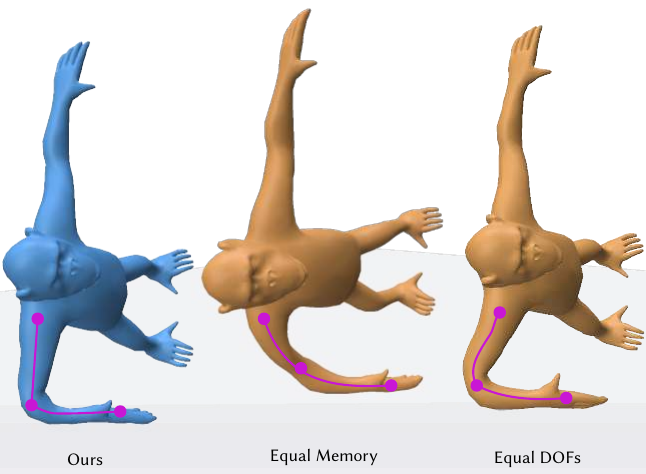}
\caption{On contact with the ground, our method correctly captures the sharp bend at the chimpanzee's elbow. At equal memory usage ($6$ handles), skinning eigenmodes noticeably smooth the deformation. Even with equal DOFs ($4.9\times$ more memory), we observe smoothing of the elbow bend. Curves are for illustration purposes only.}
\Description{Different deformations of a chimpanzee bending its arm.}
\label{fig:compare_skinning}
\end{figure}

Rather than cluster, we instead iteratively compute both handle positions and weights using an algebraic procedure analogous to mesh decimation~\citep{Garland1997}. Given $\nummodes$ target displacement vector fields, we begin with an overprescribed set of handles scattered throughout the input mesh and remove handles sequentially in a hill-descending fashion, recomputing weights at each iteration. Our decimation-based approach naturally places handles in a manner that accounts for thin geometric features and differing material properties (since they are encoded in the target displacement fields)~(\autoref{fig:dist}). Because we remove handles in error-minimizing order, we can tightly control the final fitting error between our output handle-based ROM and the target displacement fields via a user-defined tolerance. Finally, we guarantee that each vertex in the mesh will be influenced by 
only a small fixed number of handles,
%at most $\hsupp$ handles. Since $\hsupp$ is typically small (we use $6$), 
Consequently, total weight storage remains approximately equal to storing rigid modes from linear modal analysis, no matter how many handles are used.
%and therefore, the final memory storage cost for our weights is approximately the same as storing just the rigid modes from linear modal analysis, no matter how many handles are used. 
This leads to drastically higher quality animations for the same basis function storage cost~(\autoref{fig:compare_linear},~\autoref{fig:compare_skinning}).

%% file: sections/preliminaries.tex
\label{sec:prelim}

We extensively rely on handle-based deformation in the remaining sections. The handle-based deformation function is
\begin{equation}
\defx(\refx) = \sum_{\sphandle\in\handleset}\wfun_\sphandle(\refx)\Taffine_{\sphandle}{}\begin{bmatrix}(\refx-\hori_\sphandle)\\1\end{bmatrix}
\label{eq-skinning}
\end{equation}
where $\defx$ and $\refx$ are the deformed and undeformed positions, $\mathcal{H}$ is the set of handle indices, and for each handle $h \in \mathcal{H}$: $w_\sphandle(\refx)$ is the scalar weight at $\refx$, $\Taffine_\sphandle\in\mathbb{R}^{3\times4}$ is the affine transformation matrix, and $\hori_\sphandle$ is the handle's origin. The term handle is used in the common, skinning animation sense, referring to the union of the affine transform and the handle center. 
In practice, we evaluate this only at mesh vertices $v \in \vertices$, where each vertex is influenced by only a small number of handles.  Letting $S(h) \subset \vertices$ denote the vertex support of handle $h$ (\autoref{fig:overview}), we restrict the sum to $h \in S^{-1}(v)$ rather than all \(\nhandles\) handles.

Rather than using deformed positions, as in the skinning literature, we work with displacements $\mathbf{u}(\refx) = \defx (\refx) - \refx$, which yields
\begin{equation}
\mathbf{u}(\refx) = \sum_{\sphandle\in\handleset}\wfun_\sphandle(\refx)\Uaffine_{\sphandle}{}\begin{bmatrix}(\refx-\hori_\sphandle)\\1\end{bmatrix},
\label{eq-skinning_disp}
\end{equation} where $\Uaffine_h = \begin{bmatrix}\mathbf{A}_h-I && \mathbf{p}_h - \mathbf{C}_h\end{bmatrix}$, $\mathbf{A}_h$ is the $3\times3$ linear component of $\Taffine_h$ and $\mathbf{p}_h$ is the $3\times1$ translation component, i.e., $\Uaffine_h$ stores the displacement away from the reference handle state. 
%(see \autoref{sec-objective} (\autoref{eq:eq-skinning-u})).

%% file: sections/methods.tex
\begin{figure}
\begin{center}
    \includegraphics[width=\linewidth]{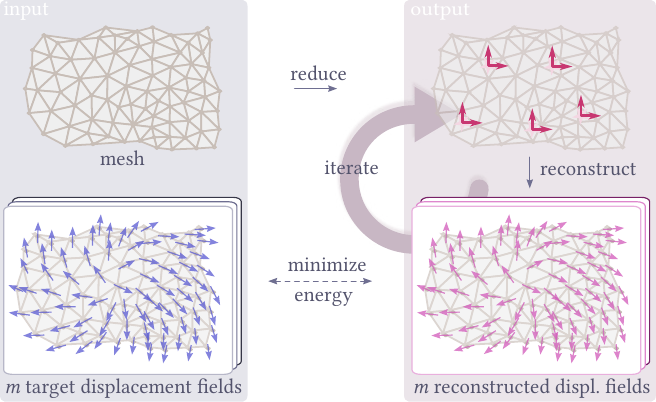}
    \caption{\label{fig:algo} Given a tetrahedral mesh and target displacement vector fields, our algorithm outputs a reduced set of handles that reconstruct the displacement vector fields by minimizing an energy (\autoref{eq:fit_energy}).}
    \Description{Diagram showing reconstructed vector fields.}
\end{center}    
\end{figure}

% PGK: small revision to move the math into 4.1 as 4.1 was almost repeating what was stated in this intro paragraph.  See original text in comments below. 
Our method takes in a high-resolution 3D tetrahedral mesh and a set of user-supplied target deformation vector fields. From these, it outputs either a user-controlled or an error-bounded set of deformation handles and weight functions. These functions are chosen in a greedy hill-descending fashion to maintain agreement with the target fields. By construction, each vertex $v$ is supported by a small fixed number of the total handles, ensuring our weight functions are sparse, compactly supported, and efficient to store.

% Our method takes in a 3D deformable solid represented by a high-resolution tetrahedral 
% mesh \(\mesh\), where \(\vertices\) is the set of
% vertices, \(\tets\) is the set of tetrahedra, and \(\refx_{v}\) is the position of vertex $v\in\vertices$.  
% Our method outputs either a user-controlled or an
% error-bounded set of deformation handles and weight functions. These functions are chosen in
% a greedy hill-descending fashion to maintain agreement with a set of
% user-supplied deformation vector fields on the mesh. By construction,
% each vertex in the tetrahedral mesh is supported by \(\hsupp \ll \nhandles \)
% deformation handles, which makes our weight functions sparse and
% compactly supported, and efficient to store.

\subsection{Problem Setup}\label{sec-thesetup}

Let the input solid be represented by a tetrahedral mesh $\{\vertices, \tets\}$, where $\vertices$ is the set of vertices, $\tets$ is the set of tetrahedra, and $X_v$ is the position of vertex $v \in \vertices$. Alongside the mesh, we require a per vertex finite family of target displacement vector fields $\{\fieldv{m}_v\}_{m=0}^{k-1}$, where $\fieldv{m}_v \in \mathbb{R}^3$ is the $m^{th}$-field value at vertex $v$. In this paper, we use the linear vibration modes of the input mesh as these fields.

% In addition to a tetrahedral input mesh, we require the user to provide a
% finite family of \emph{target displacement 
% %vector % PGK: word vector makes this wordy... clear it is a vector field, no?
% fields}
% \(\{\fieldv{m}\}_{m=0}^{\nummodes-1}\), where
% \(\fieldv{m}_v \in \mathbb{R}^3\) is the field value at vertex $v$. In this paper, we use linear vibration modes of the input tetrahedral mesh as these fields.

We initialize our method with a dense set of deformation handles $\handleset$, placing one handle $h \in \handleset$ at the centroid $\hori_h$ of each tetrahedron $t \in \tets$ (so $\vert{}\handleset\vert{} = \vert{}\tets\vert{}$ at setup). 
Additionally, we compute $\fieldv{m}_h$ for each $t \in \tets$ (and handle $h$) as the average of $\fieldv{m}_v$ at the vertices of $t$.
For each target displacement field, we define the handle displacement transformation
$\Uaffine_h^m = \begin{bmatrix} (\grad{\fieldv{m}})_h & \fieldv{m}_h \end{bmatrix}$,
where $(\grad{\fieldv{m}})_h$ is the constant piecewise-linear gradient evaluated over tetrahedra $\tets$. These handle transformations reproduce the affine motion of their respective tetrahedra prior to weight computation. 

% For each target displacement field $\fieldv{m}$, the initial handle transformation $T_h^m$ is given by 
% $T_h^m=\begin{bmatrix}\grad{\fieldv{m}_{h}} &\fieldv{m}_h\end{bmatrix}$.
% Here, gradients are computed via the piecewise linear gradient operator defined on the tetrahedra $\tets$ for the linear part of this transformation, while the translation part is the centroid's displacement. 
% %
% Applying these handle transformations as a skinning transform on each tetrahedron's vertices will exactly match the target displacement fields. Although this makes initial weight assignment trivial, the overall objective is to approximate target displacement fields using a significantly reduced set of handles.

%i.e., linear part of each displacement skinning transform is the spatial gradient of the $m^{th}$ vector field at the $c^{th}$ centroid and the translation portion is the displacement, $\fieldv{m}_c$.

We remove handles one at a time via a hill-descending algorithm until either a
user-defined error threshold \(\errgth\) or a user-defined minimum number
of handles,  $H_{\min}$, is reached (bounded below by \(\hsupp\)). 

\subsection{Decimation Overview}\label{sec-overview}

Before starting decimation, we first compute the small number of handles \(\hsupp \ll |\handleset|\) (we use $H=6$), 
that support each vertex (thus, $|S^{-1}(v)|=6$). 
This is done using Yen's algorithm~\citep{yen1970algorithm}, a relative of the Bellman-Ford algorithm~\citep{bellman1958routing}, which is especially amenable to parallelization. Given these supports and a reconstruction objective (\autoref{sec-objective}), our algorithm removes handles from the mesh one at a time using a priority queue, \(\pqueue\), in which each handle is ordered
by the delta in error, \(\errdelta{}\), tied to its removal (\autoref{alg:overview}). Supports are recomputed for every vertex affected by the removal, and the corresponding
\(\errdelta{}\) values are updated~(\autoref{fig:algo}). 
%\Paul{ref fig? redo fig not as wrap? or alter margins... I can do this.}

\begin{algorithm}[t]
\SetAlgoLined
\caption{Modal Handle Decimation}
\label{alg:overview}
\KwIn{Tetrahedral mesh $\{\vertices, \tets\}$; fields $\{\fieldv{m}_v\}_{m=0}^{k-1}$; handles support $\hsupp$; threshold $\errgth$; minimum handles $H_{\min}$}
\KwOut{Active handles $\handleset$, per-vertex weights $\wv\in\mathbb{R}^{\numverts \times \nhandles}$, supports $\supp{}$}
Build per-tet, per-field affine deformations $\Uaffine_{h}^{m}$ \tcp*{ \autoref{sec-thesetup}}%
Run Yen's algorithm to obtain $\supp{}$ for all vertices $v$\ \tcp*{ \autoref{sec-overview}}%
%= [\Fblock_{jm} \mid \tblock_{jm}]$ for all tetrahedra, $j$, and all vector fields $m$\;
Build augmented volume graph $\volgraph$ \tcp*{\autoref{fig:distance_graph}}
Compute smooth compactly-supported initial weight functions \tcp*{\autoref{sec:initial_weights}}
Build priority queue $\pqueue$ with per-handle trial costs $\errdelta{j}$\ \tcp*{\autoref{sec:decimation-loop}}%
\While{$\errg < \errgth \;\mbox{or}\; |\handleset| > H_{\min}$}{
  Pop $\handle^*$ with smallest $\errdelta{j}$ from $\pqueue$\;
  Remove $\handle^*$ from active supports\;
  Run Yen's to update $\supp{v}$ for 
  %all  % PAUL: supressing this word to avoid the line break 
  affected vertices\tcp*{\autoref{sec-volume-graph}}
  Update weights $\wv_v$ for all affected vertices\tcp*{\autoref{sec:update}}
  Update $\errg$; refresh dirty $\errdelta{j}$ in $\pqueue$\ \tcp*{\autoref{sec:decimation-loop}}%
}
\Return $\{\handleset,\wv, \supp{}\}$\;
\end{algorithm} 

% \begin{algorithm}[t]
% \SetAlgoLined
% \caption{Modal Handle Decimation}
% \label{alg:overview}
% \KwIn{Tetrahedral mesh; fields $\{\fieldv{m}_v\}_{m=0}^{k-1}$, initial weights $\wv\in\mathbb{R}^{\numverts \times \nhandles}$; handles $\hsupp$; threshold $\tau$; minimum handles $H_{\min}$}
% \KwOut{Active handles $\handleset$, per-vertex weights $\wv\in\mathbb{R}^{\numverts \times \nhandles}$, supports $\supp{}$}
% Build per-tet, per-field affine deformations $\Uaffine_{h}^{m}$ \tcp*{ \autoref{sec-thesetup}}%
% %= [\Fblock_{jm} \mid \tblock_{jm}]$ for all tetrahedra, $j$, and all vector fields $m$\;
% Build augmented volume graph $\volgraph$ \tcp*{\autoref{fig:distance_graph}}
% Run Yen's algorithm to obtain $\supp{}$ for all vertices $v$\ \tcp*{ \autoref{sec-volume-graph}}%
% Build priority queue $\pqueue$ with per-handle trial costs $\errdelta{j}$\ \tcp*{\autoref{sec:decimation-loop}}%
% \While{$\errg < \tau \;\mbox{and}\; |\handleset| > H_{\min}$}{
%   Pop $\handle^*$ with smallest $\errdelta{j}$ from $\pqueue$\;
%   Remove $\handle^*$ from active supports\;
%   Run Yen's to update $\supp{v}$ for all affected vertices\;
%   Update weights $\wv_v$ for all affected vertices\;
%   Update $\errg$; refresh dirty $\errdelta{j}$ in $\pqueue$\ \tcp*{\autoref{sec:decimation-loop}}%
% }
% \Return $\{\handleset,\wv, \supp{}\}$\;
% \end{algorithm} 

\subsection{Vector-Field Fitting Objective}
\label{sec-objective}
Given a current set of handles, \(\handleset\), their transformations, $\Uaffine$, positions $\hori$ and a set of per-vertex
deformation weights \(\wv\), we characterize their optimality by the
difference between the target displacement fields, $\{\fieldv{m}_v\}_{m=0}^{k-1}$, and their
reconstruction $\{\tilde{\mathbf{u}}_v^m\}_{m=0}^{k-1}$ using the current handle and weight set.
While \(\wv\) has dimension $\numverts\times\nhandles$, our per-vertex handle limit, $\hsupp$, allows us to store it as a sparse matrix with total storage cost \(\numverts\times \hsupp\) in a compressed row
format.  Recall, \(\supp{v}\) stores the \(\hsupp\) indices of the handles
supporting vertex \(v\), 
% PGK this reminder of \supp could be suprressed... only useful if you want to think of it as being part of the matrix datastructure, i.e., specifically CSR.  I note that likewise \supp is not used in the displacement equation (2) below... but it COULD BE
%and \(\wvc \in \mathbb{R}^{\numverts\times\hsupp}\) stores the
%corresponding weights for each indexed handle;
 and $\mathbf{\wfun}_v$ is the vector of weights associated with vertex $v$ (that is, the $v^\text{th}$ row of $\wv$).

We define the reconstructed target field
relative to the vertex's weight vector $\mathbf{\wfun}_v$ for any given subset of handles, \(\handleset\),
using \autoref{eq-skinning_disp} in its fully discretized form as 
\begin{equation}
\tilde{\mathbf{u}}_v^m(\mathbf{\wfun}_v) = \sum_{\disphandle\in\supp{v}}\wfun_{\sphandle v}\Uaffine_{\sphandle}{}\begin{bmatrix}(\refx_v-\hori_\sphandle)\\1\end{bmatrix},
\label{eq-skinning-u}
\end{equation}
%
% \begin{align}
% % PGK: again, this label definition makes the equation number be used as a section number.  :( 
% %\protect\phantomsection
% \label{eq:eq-skinning-u}
% \tilde{\fieldv{}}^m_v(\mathbf{\wfun}_v)  %\sum_{\disphandle\in\handleset}
% % %{c=0}^{\nhandles-1}
% % \wfun_{v\disphandle}
% % %\begin{bmatrix}\grad{\fieldv{m}_{\disphandle}} &\fieldv{m}_\disphandle\end{bmatrix}
% % \: (T_h^m - T_{h0})
% % \begin{bmatrix}\refx_v-\hori_\disphandle\\1\end{bmatrix},\\
% &= \left( 
% %\sum_{\disphandle\in\handleset}
% \sum_{\disphandle\in\supp{v}}
% \wfun_{v\disphandle} \: T_h^m
% \begin{bmatrix}\refx_v-\hori_\disphandle\\1\end{bmatrix}
% \right) - \refx_v
% \end{align}
% % PGK: would be nice to have reader see mathcal H as a subset of all handles, and use index h rather than c, having preivously defined a handle at each centroid.
where $m$ indexes the target displacement field, $v$ indexes mesh vertices, and
$\refx_v$ is the undeformed position of vertex $v$. 
%$T_{h0}=\begin{bmatrix}I&\hori_h\end{bmatrix}$
%$\hori_\disphandle$ is the undeformed position 
%is the undeformed pose
%of handle $\disphandle$, 
% and the vertex's weight vector $\mathbf{\wfun}_v$ has components $\wfun_{v\disphandle}$.
Given the fixed set of handle transformations, these weights uniquely determine the reconstructed displacements. 
%\paul{(Updated: @Dave, correct? Also, still need to redefine $\refx_v$?}

%for handle/centroid $\disphandle$ and 
%$\wfun_{v\disphandle} = \wfun_\disphandle(\refx_v)$. 
%We set the linear part of each displacement skinning transform to be the spatial gradient of the $m^{th}$ vector field at the $c^{th}$ centroid and the translation portion to be the displacement, $\fieldv{m}_c$. 
%
%Gradients are computed using the piecewise linear gradient operator defined on the tetrahedra $\tets$. Using $\begin{bmatrix}\grad{\fieldv{m}_{c}} &\fieldv{m}_c\end{bmatrix}$ as the per-handle transform ensures that the computed weights reconstruct target vector fields when the per-tetrahedral handles match the local deformations prescribed by said vector fields.

Decimation minimizes a normalized reconstruction error. For a given set of weights, we define the per-vertex error
\begin{equation}
\errv{v}(\mathbf{\wfun}_v) = \frac{1}{\nummodes}\sum_{m=0}^{\nummodes-1}
\frac{\rho_v\,\|\tilde{\fieldv{}}^m_v(\mathbf{\wfun}_v) - \fieldv{m}_v\|_2}
     {\mscale{m}},
\label{eq:per_vertex_error}
\end{equation}
where $\rho_v$ is the lumped FEM mass associated with vertex $v$, and
the mass-weighted mode norm is
\begin{equation}
\mscale{m} = \left(\sum_{v\in\vertices} \rho_v \|\fieldv{m}_v\|_2^2\right)^{1/2}.
\label{eq:mode_scale}
\end{equation}
For mass-normalized vectors, such as the generalized eigenvectors (\autoref{sec:rom}), $\mscale{m} = 1$.
The global reconstruction error is then the root-mean-square of these per-vertex errors
\begin{equation}
\errg(\wv) = 
%\frac{1}{\numverts}
\left( \sum_{v\in\vertices}
\errv{v}^2(\mathbf{\wfun}_v) \right)^{1/2}.
\label{eq:global_error}
\end{equation}
%This $\errg$ is what the stopping criterion and removal ranking use (\autoref{sec:decimation-loop}). We use this normalized metric because it makes the decimation error dimensionless and therefore allows a user to specify a relative, unitless tolerance which is far more intuitive and makes selected thresholds transferable across different geometries.
%
% PGK: add an expliti min... also vague with "we can" as Section 4.4 changes the objective
For a given set of handles, we can compute the optimal weights by explicitly minimizing this global error
\begin{equation}
\wv^* = \arg\min_{\wv} \errg(\wv),
\label{eq:weight_minimization}
\end{equation}
where optimization is over the nonzero elements of the sparse matrix $\wv$.
% (Optional: Add "subject to \sum \wfun{vk} = 1, \wfun_{vk} \ge 0" here if standard LBS constraints apply)
The resulting minimized error $\errg(\wv^*)$ serves a dual purpose: it provides the metric used to rank handles for removal, and acts as the final stopping criterion for the decimation loop (\autoref{sec:decimation-loop}). We use this specific normalized formulation because it renders the decimation error dimensionless. This allows the user to specify a relative, unitless tolerance, which is far more intuitive and ensures that selected thresholds remain directly transferable across different geometries.

\subsection{Initial Setup}
\label{sec:initial_weights}

Before decimation can begin, we require smooth and compactly-supported weight functions that satisfy partition of unity for
our dense initial set of handles. For all weight computations, we use the following quadratic approximation (rather than minimize the normalized objective \autoref{eq:global_error} directly) to yield easy-to-solve quadratic minimizations: 
\begin{equation}
E(\wv) = \sum_{m=0}^{\nummodes-1}\sum_{v\in\vertices} \rho_v \| \tilde{\fieldv{}}^m_v(\mathbf{\wfun}_v) - \fieldv{m}_v \|^2_2.
\label{eq:fit_energy}
\end{equation} Unlike \autoref{eq:per_vertex_error}, this energy sums \emph{squared}, mass-weighted residuals and is therefore quadratic in the weights.

We compute the initial weights by minimizing a biharmonic-regularized fit to the input target displacement fields,
\begin{equation}
E_{\text{initial}}(\wv) =  E(\wv) + \lambda\cdot
  \sum_{h\in\handleset}
  %=0}^{\nhandles-1}
  \selectW^T\wv^T 
  B_h
  %\selectV^T B \selectV 
  \wv\selectW, %PGK was B_h in the middle originally... could be tighter and fixed with def below.
\end{equation}
%\paul{drop the h subscript on B and modify the $B_h$ description below, or change the SHTBHSH sandwich to be just $B_h$ and use the definition below! }
where \(E(\wv)\) is the least-squares energy~(\autoref{eq:fit_energy}) measuring agreement between the reconstructed and target displacement fields, and 
$\selectW\in\mathbb{R}^{\nhandles\times1}$ selects the column of $\wv$ corresponding to handle $h$.
Here, $B_h = \selectV^T \selectV B \selectV^T\selectV \in \mathbb{R}^{|\vertices|\times|\vertices|}$ is a sparse, per-region discrete biharmonic operator obtained by restricting the tetrahedral mesh's global biharmonic operator, $B\in\mathbb{R}^{|\vertices|\times|\vertices|}$, 
using $\selectV \in \mathbb{R}^{|\invsupp{h}|\times|\vertices|}$ to select the vertices in $\invsupp{h}$ (i.e., treating weights outside that region as zero). The resulting sparse, partition-of-unity weights seed the decimation procedure described next.

% This was a sub sub sub... but not part of initial setup yes??
\subsection{Efficient Weight Update}
\label{sec:update}
Our algorithm removes a single handle at a time from the input geometry and recomputes the weight functions at all affected vertices. Performing this update independently for all vertices makes maintaining smoothness and the partition-of-unity constraint expensive. Adding regularizers and constraints couples the problem, yielding linear solves with size on the order of the number of vertices affected by a removed handle. We avoid both issues by not recomputing weights from scratch after handle removal; instead, we redistribute the weight function associated with the removed handle $\handle^*$, to its local support.
% neighborhood. \paul{to its support? neighorhood make me think vertex adjacency. }

\begin{figure}[t]
\centering
\includegraphics[width=\columnwidth]{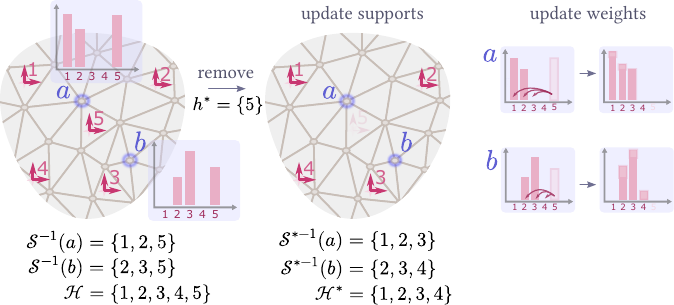}
\caption{When a handle, say $h^* = 5$, is removed, support sets and their inverses are updated. Then, at each vertex, weights from the removed handle are redistributed to the handles in the updated inverse support set while maintaining partition of unity.}
\Description{Diagram showing handle updates and weight redistribution to new handle sets.}
\label{fig:handle_removal}
\end{figure}

Prior to removal we find the subset of vertices $\invsupp{\handle^*}$ that are supported by $\handle^*$~(\autoref{fig:handle_removal}). We then compute new handle support sets $\newsupp{v}$ for all vertices $v$, affected by the impending handle removal (i.e., $\handle^*$ removed and another handle added for each vertex, see \autoref{sec:support} and \autoref{sec:decimation-loop}). We then redistribute the weight function associated with $\handle^*$ within a new local handle set defined by 
\begin{equation}
\mathcal{H}^*=\underset{v\in\invsupp{\handle^*}}{\bigcup}\newsupp{v}.
\label{eq:localhandleset}
\end{equation}
% PGK: cup -> bigcup and $$ instead of $ to have unumbered equation...
% or revert, but it didn't look too pretty inline :/
By construction the new support set of each vertex only contains $\hsupp$ handles and so the updated handle discretization will remain compactly supported. Despite the fact $\mathcal{H}^*$ is a union of each affected vertex's supporting handles, the compact support leads to $\mathcal{H}^*$ being small (\autoref{fig:handle_removal}). In our experiments we use $\hsupp=6$ and observe $|\mathcal{H}^*|\leq 10$ in all cases.

Again, our removal is built around~\autoref{eq:fit_energy} which yields a small, easy-to-solve KKT system after the reparameterization below---important because a single decimation run may perform millions of weight recomputations. Ranking and stopping continue to use the normalized objective \autoref{eq:global_error} evaluated after each update.

Concretely, for each $v\in\invsupp{\handle^*}$ and each handle $h\in\mathcal{H}^*$, we compute updated weight functions 
\begin{align}
\wfun^*_{vh} = \wfun_{vh} + \alpha_h \wfun_{v h^*}.
\label{eq:weight_update}
\end{align}
We seek the weight redistribution vector $\boldsymbol{\alpha} = (\alpha_h)_{h \in \mathcal{H}^*}$ that minimizes the fit energy.  Substituting the updated weights $\mathbf{w}_v^*(\boldsymbol{\alpha})$ defined by \autoref{eq:weight_update} into \autoref{eq:fit_energy} and restricting the sum to affected vertices yields 
%Substituting the updated weights $\mathbf{w}_v^*(\boldsymbol{\alpha})$ defined by \autoref{eq:weight_update} into \autoref{eq:fit_energy} and restricting the sum to affected vertices, we obtain the optimization problem over $\boldsymbol{\alpha} = (\alpha_h)_{h \in \mathcal{H}^*}$ with energy
%Substituting into \autoref{eq:fit_energy} and restricting the outer sum to affected vertices, our optimization problem, now expressed over $\boldsymbol{\alpha} = (\alpha_h)_{h \in \mathcal{H}^*}$, with new weights $\mathbf{\wfun}_v^*(\boldsymbol{\alpha})$ defined by \autoref{eq:weight_update}, uses
\begin{equation}
E_\text{update}(\boldsymbol{\alpha}) = 
\sum_{m=0}^{\nummodes-1} \sum_{v\in\invsupp{\handle^*}} \rho_v 
\| \tilde{\fieldv{}}^m_v( \mathbf{\wfun}_v^*(\boldsymbol{\alpha})) - \fieldv{m}_v\|^2_2.
\label{eq:update_solve}
\end{equation}
%
% \begin{equation}
% E(\boldsymbol{\alpha}) = \sum_{m=0}^{\nummodes-1}\sum_{v\in\invsupp{\handle^*}}\rho_v\|\tilde{\fieldv{}}^m_v(\mathbf{\wfun}_v + \wfun_{v h^*}P_v\boldsymbol{\alpha}) - \fieldv{m}_v\|^2_2,
% \label{eq:update_solve}
% \end{equation}
%where $P_v\in \mathbb{R}^{\hsupp\times|\boldsymbol{\alpha}|}$ selects the components of $\boldsymbol{\alpha}$ associated with the handles supporting $v$. 
Because this quadratic objective is minimized with respect to $\boldsymbol{\alpha}$, the number of variables 
$|\mathcal{H}^*|$
%$|\boldsymbol{\alpha}|$ 
will typically (except for the first few iterations) be much smaller than the number of vertices in $\invsupp{\handle^*}$, and because the resulting weight functions are all linear combinations of those from the previous state, they maintain all smoothness properties since they span a subset of that previous function space. 

% PGK: smoothed this over... would read better without all the mentions of poo
To enforce partition of unity (\POU),
%a partition of unity (\POU),
we must constrain the minimization such that 
$\sum_{h\in\newsupp{v}} \wfun^*_{vh}=1$ for every vertex.
%\sum_{h=0}^{\hsupp-1} (\wfun_{vh}+\wfun_{v h^*}\alpha_h) = 1$.
Since the initial weight functions already sum to one, this property is preserved provided 
$\sum_{h\in\newsupp{v}}\alpha_h = 1$~(\autoref{sec:pou}). 
%$\sum_{h=0}^{\hsupp-1}\alpha_h = 1$~(\autoref{sec:pou}). 
% PGK: this indexing (and the sum in the first sentence too) was not clear as $\alpha$ has a size bigger than $H$ 
%
%In order to ensure \POU we must constrain the above minimization so that $\sum_{h=0}^{\hsupp-1} (\wfun_{vh}+\wfun_{v h^*}\alpha_h) = 1$. Because the weight functions prior to the update possessed \POU, \POU is preserved if $\sum_{h=0}^{\hsupp-1}\alpha_h = 1$~(\autoref{sec:pou}). 
%
Applying this constraint naively per-vertex would lead to a large, ill-conditioned optimization problem. The solution is to leverage the observation that each vertex is supported by exactly $\hsupp$ handles and 
% PGK: skip this part of the sentence... and GET TO THE POINT... shared combinations!
%thus is affected by at most $\hsupp$ values in $\boldsymbol{\alpha}$. Therefore, many 
that many vertices share the same combination of $\hsupp$ values from $\boldsymbol{\alpha}$. Rather than applying the \POU constraint per-vertex, 
we instead collect all distinct handle subsets present across the vertices $v\in\invsupp{\handle^*}$ into a family 
%$\mathcal{F}$.
$\mathcal{F} \subset \mathcal{P}(\mathcal{H^*})$ where $\mathcal{P}(\mathcal{H^*})$ denotes the power set of the local handle set.
We then apply a constraint per unique combination of handles,
%instead apply it per-unique combination of $\boldsymbol{\alpha}$ values, 
which significantly reduces the size of the constraint matrix in later iterations of decimation when vertex supports become large. This leads to a linearly constrained quadratic optimization problem that computes optimal weight updates and satisfies \POU but does not scale with vertex patch size
\begin{align}
\boldsymbol{\alpha}^* &= \arg\min_{\boldsymbol{\alpha}} E_\text{update}( \boldsymbol{\alpha} ) \\
\text{s.t.} \quad & \sum_{h \in \mathcal{H}_i} \alpha_h = 1, \quad \forall \: \mathcal{H}_i \in \mathcal{F}.
\end{align}

%will answer later :)
%\paul{so how many constraints are typically required? i.e., size of $|\mathcal{F}|$? how are unique combinations efficiently obtained?} 

Importantly, despite all the constraints, this optimization problem always admits a solution since setting $\alpha_i=\frac{1}{\hsupp}$ uniformly satisfies all \POU constraints. Despite this property, in practice, this minimization can become ill-posed, especially when vertex supports are small at the beginning of the decimation loop. To tackle this we apply cascading regularization to the problem (in the spirit of \citet{osqp}). We first attempt to solve the unregularized problem and if this fails we try again with regularization weights $10^{-7}$, followed by $10^{-4}$. Since we diagonally regularize the whole KKT system~\citep{osqp}, this induces small constraint violations which we project out explicitly. If regularization fails, we fall back to $\alpha_i=\frac{1}{\hsupp}$ as the solution. Experimentally, this progressive approach has proven to be robust and yields visually high-quality results.

\subsection{Efficient Vertex Support Identification}
\label{sec:support}\label{sec-volume-graph}
\begin{figure}[t]
% \centering
\includegraphics[width=0.92\linewidth]{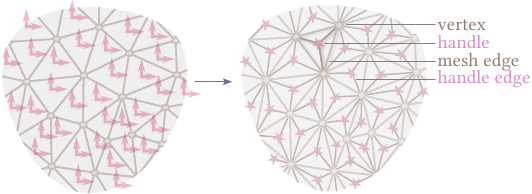}
\caption{\emph{Left:} We initialize our method with a dense set of handles (one per tetrahedral centroid). \emph{Right:} We build an augmented graph to be used for handle-vertex and handle-handle distance computation. The augmented graph's vertex set includes  mesh vertices and handles and its edge set includes mesh edges and `handle edges' that  connect mesh vertices to handle positions.}
\Description{Diagram showing the extended graph used by Yen's algorithm.}
\label{fig:distance_graph}
\end{figure}

Handle decimation requires repeatedly computing the $\hsupp$ nearest active handles that support any given vertex to update $\supp$. This is a $k$-nearest-handles shortest-path problem, requiring us to compute shape-aware shortest paths from each vertex to its nearest $\hsupp$ handles. We solve this problem using the well-known all-pairs shortest-paths modification to the iterative Bellman-Ford algorithm~\citep{bellman1958routing}, known as Yen's algorithm~\citep{yen1970algorithm}. We execute the algorithm on a modified mesh with nodes inserted at tetrahedral centroids to represent handle positions, and with new edges to connect the centroids to the existing mesh vertices~(\autoref{fig:distance_graph}). 
%Rather than store distances to all handles at each vertex, we store the $\hsupp$ nearest along with one extra handle, so that $\hsupp+1$ handles are cached. When a handle is removed, we can evaluate the trial removal by filtering out the removed handle and using the remaining $\hsupp$ nearest handles from the cached set, avoiding an additional graph query. 
We terminate the iterations when the distance lists at all vertices reach stationary points. We chose this iterative approach for two reasons. First, upon handle removal, we can warm-start the algorithm with its previous settings, leading to fast convergence. Second, the algorithm is highly amenable to GPU acceleration. Every vertex support recomputation is followed by a 
% inverse 
support map $\invsupp{}$ update, which is stored in a simple strided linear array.  %Yes, fixed in first sentence\paul{are both $\supp$ and $\invsupp$ updated?}

%\begin{wrapfigure}{r}{0.5\linewidth}
%\end{wrapfigure}
% PGK: this figure seems most useful here rather than where it is earlier... Also is too wide to work well as a wrap fig, as well as having a caption that was too long.

\subsection{Decimation Loop}\label{sec:decimation-loop}

After computing the initial weights~(\autoref{sec:initial_weights}), we populate the priority queue one handle at a time. For each initial handle (one per tetrahedron centroid in the input mesh), we remove that handle, recompute the support and weights using our efficient update~(\autoref{sec:update}), and compute the removal cost $\errdelta{j}$ defined in \autoref{eq:delta_error} below.

The decimation orders handles by the change in the normalized global fitting objective~(\autoref{eq:global_error}). For each active handle $j$, let $\wv^{j}$ be the weights obtained after a trial removal (approximate weight update of \autoref{sec:update}, then evaluation of \autoref{eq:global_error}). The priority-queue key is the resulting change in this global error:
\begin{equation}
\errdelta{j} = \errg(\wv^{j}) - \errg(\wv).
\label{eq:delta_error}
\end{equation}
After accepting the removal of $j$, the global error becomes $\errg \gets \errg + \errdelta{j} = \errg(\wv^{j})$.

Once the priority queue is filled we proceed to continually remove the remaining handle with minimum error until reaching either a prescribed error tolerance, $\errgth$,
%\TS{correct?}, 
or a minimum number of remaining handles in the object, $H_{\min}$. After every removal, we must update the error stored in the priority queue for any handle which is influenced by a vertex in $\invsupp{\handle^*}$, which we do by performing a trial removal for each {\em dirty} handle,
% PGK: put dirty in em rather than scare quotes (i.e., bit of a  definition?) and removed quotes in later use of word dirty
recomputing the weights using the method in~\autoref{sec:update}, computing this new error, updating the priority queue and replacing the handle. We make a small optimization 
%to support computation  # PGK could read as "in support of" 
for the computation of supports~(\autoref{sec:support}): instead of computing the $\hsupp$ nearest handles to each vertex, we compute the $\hsupp+1$ handles. This way, when we perform updates on dirty handles we do not need to rerun Yen's algorithm each time; rather, we simply ignore the dirty handle and use this additional nearest handle in its place. 

\subsection{Hyper-Reduced Handle-Based Simulation}

Once we have computed a small set of handles for an object we can perform fast elastodynamics simulations using the standard formulation. Our simulation algorithm is not a contribution and is designed only to demonstrate the suitability of our output handle sets for visually compelling, real-time elastodynamics simulations. Because of this, we rely on a first-order, position-level, linearly implicit time integration scheme~\citep{baraff2023large} with penalty springs for contact handling and no friction model. After reduction, the bottleneck in nonlinear reduced-order simulations is almost always integrating energies, gradients, and Hessians to perform the implicit position-update calculation, which we alleviate using cubature~\citep{an2008optimizing, vonTycowicz2013}.

\begin{figure}[t]
\centering
\includegraphics[width=\columnwidth]{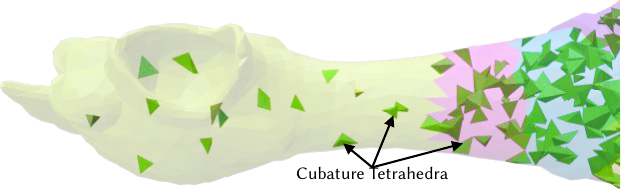}
\caption{Compactly-supported weights allow us to decompose an input tetrahedral mesh into large elements, seen here in different colors. For each element we compute a reduced set of cubature tetrahedra (green) for fast runtime integration of energies, gradients, and Hessians.}
\Description{Transparent rubber chicken head showing elements and cubature tetrahedra.}
\label{fig:cubature_elements}
\end{figure}

Our cubature scheme is compact-support-aware and does not rely on our input target displacement vector fields. The compact support of our handles allows us to divide our object into {\em elements} (\autoref{fig:cubature_elements}), 
defined as collections of tetrahedra where at least one vertex is influenced by a given distinct handle subset from our family $\mathcal{F}$ (\autoref{sec:update}).
That is, for a unique handle subset $\handleset_i\in\mathcal{F}$, 
we define an element $E$ by its constituent set of tetrahedra.  Because tetrahedra along the element boundary are shared and share vertices with adjacent elements, we also define 
$\handleset_E \supseteq \handleset_i$ as the complete set of handles that influence the tetrahedra of the element, i.e., the union of $\supp{v}$ for all vertices of all tetrahedra in $\tets_E$ (see also \autoref{eq:localhandleset}, and note that we observe $|\handleset_E|$ to be less than 10 in practice).

% PGK: DON'T need to mention the overlap where tets on teh boundary belong to both... Would stating this make it more confusing, or help??
%
%which we define as collections of tetrahedra with at least one vertex influenced by an identical set of $\hsupp$ handles.

We compute a separate set of cubature tetrahedra for each element using an iterative procedure. This is necessary because even single elements can include more tetrahedra than can efficiently be processed by high-performance NNLS codes (for instance, our Rose example features a single element with over $50{,}000$ tetrahedra).
Overall, the process we use to compute cubature weights is as follows:
(1) start with a given set of candidate tetrahedra from within the element;
(2) solve the NNLS problem for cubature tetrahedra;
(3) if the cubature error is above threshold, add more candidate tetrahedra and repeat step (2).
% \begin{enumerate}
%     \item Start with a given set of candidate tetrahedra from within the element;
%     \item Solve the NNLS problem for cubature tets;
%     \item If the cubature error is above threshold, add more candidate tetrahedra and repeat step (2).
% \end{enumerate}

\subsubsection{Selecting Candidate Tetrahedra}

Given a vector $\mathbf{f}$, we define its spectral order $k$ as the index of the smallest generalized eigenvalue of the Laplacian, $L$, that is greater than $\frac{\mathbf{f}^TL\mathbf{f}}{\mathbf{f}^TM\mathbf{f}}$, which is the spectral energy (Rayleigh quotient) of $\mathbf{f}$~\citep{dong2024rayleighquotientgraphneural}. Here $L$ (resp. $M$) is the Laplacian (resp. consistent mass matrix) for the tetrahedra making up an element of our handle discretization. It is well known from the theory of cubature that the number of quadrature points required to accurately integrate a function is proportional to the spatial frequency of that function; thus, the higher the spectral energy, the more cubature samples we require. Inspired by this, we randomly initialize our candidate tetrahedron sets $\mathcal{T}_E^c \subset \mathcal{T}_E$ with $2k$ elements and likewise expand them in increments of $2k$. While there is no exact relationship to draw on, because our method incrementally adds $2k$ candidate tetrahedra each iteration, it will eventually satisfy a user-set error tolerance ($1e^{-4}$). Because NNLS solves are sparsifying~\citep{an2008optimizing,vonTycowicz2013}, having extra tetrahedra in the candidate set is acceptable.
% \paul{this section is ok, but starts in an abstract dry way! provides a lot of info about justifying 2k tets, but omits saying how selected.  Adding the word "random" perhaps would help... otherwise other papers (von Tycowicz) I believe has some ranking on which tets are likely to be valuable!}

\subsubsection{Calculating cubature weights} 
Once we have a candidate tetrahedron set, $\tets_E^c$, we solve for per-element cubature weights, guided by two rules-of-thumb. First, we want our optimized cubature rule to accurately integrate the per-element shape functions, defined by the skinning weights in the element domain. Second, we want to accurately integrate elastic forces in the element. 

For each handle, $h$, in an element we define its basis function as the product of its associated weight with the linear polynomial basis from the handle itself: 
\begin{equation}
    \phi_h(\mathbf{X})= \wfun_{h}\left(\mathbf{X}\right)\begin{bmatrix}1&X&Y&Z\end{bmatrix}^T.
\end{equation}

Cubature, amounts to approximating the exact integrals of these quantities over the whole object via a summation over a subset of per-tetrahedron integrals. Given $t\in\tets_E$, we define the per tetrahedron integrals as 
\begin{equation}
A_{ht} = \int_{t} \phi_h\left(\mathbf{X}\right)d\mathbf{X},
\end{equation} 
where $h\in\handleset_E$ and $A_{ht}\in \mathbb{R}^{4\times1}$.
%and $\handleset_E$ is the set of handles that influence the current element. 
Total integrals over the element are then 
\begin{equation}
\mathbf{b}_{h} = \sum_{t \in \tets_E} A_{ht}.
\end{equation}

For forces, we generate $p$ random perturbations of the handles in $\handleset_E$ and then compute the forces using the assigned elastic constitutive model.  For each random sample $l$ and each tetrahedron $t\in\tets_E$, we define column vector
\begin{equation}
F_{lt} = -\int_t\nabla_\mathbf{q} \psi\left(\mathbf{q}^l\right)d\mathbf{X},
\end{equation} where $\mathbf{q}^l$ is a stacked vector of handle degrees of freedom with the $l^{th}$ perturbation applied, and $\psi$ is the hyperelastic energy density function for $t$. The exact 
%\paul{only for $t \in \tets_E$? should this not be all tetrahedra for exact?} 
integral of these forces  over the element is then
\begin{equation}
\mathbf{g}_{l} = \sum_{t \in \tets_E} F_{lt}.
\end{equation}

%Following \citet{an2008optimizing}, 
We then solve a nonlinear least squares problem to arrive at a set of cubature weights, 
\begin{equation}
\boldsymbol{\omega}^* =
\arg \underset{\boldsymbol{\omega} \ge 0 }{\min} 
\left\| F \boldsymbol{\omega} - \mathbf{g} \right\|_2^2 +
\lambda \left\| A \boldsymbol{\omega} - \mathbf{b} \right\|_2^2,
% \left|\begin{bmatrix} A \\ F \end{bmatrix} \boldsymbol{\omega} - \begin{bmatrix}\mathbf{b} \\ \mathbf{g}\end{bmatrix}\right|_2^2,
\end{equation}
where $\lambda$ balances the
force integration error and the kinematic shape function integration error (we use $\lambda=1$), and the matrices $A$ and $F$ are defined block-wise,
\begin{align}
A &= \begin{bmatrix} A_{ht} \end{bmatrix}_{h \in \mathcal{H}_E, \, t \in \mathcal{T}_E^c}, \\
F &= \begin{bmatrix} F_{lt} \end{bmatrix}_{l \in \{0, \dots, p-1\}, \, t \in \mathcal{T}_E^c},
\end{align}
and similarly the block-wise vectors, 
\begin{align}
\mathbf{b} &= \begin{bmatrix} \mathbf{b}_h \end{bmatrix}_{h \in \mathcal{H}_E}, \\
\mathbf{g} &= \begin{bmatrix} \mathbf{g}_l \end{bmatrix}_{l \in \{0, \dots, p-1\}}.
\end{align}
Note that we do not perform the relative-error normalization of~\cite{an2008optimizing} as we obtain stable results for large non-linear deformations without it.

%% file: sections/results.tex
All stages of our algorithm are implemented in NVIDIA Warp~\citep{warp2022} except for the initial weight computation, which is implemented in Python using SciPy~\citep{2020SciPy-NMeth} and libigl~\citep{libigl} to compute the biharmonic operator~(\autoref{sec:initial_weights}), as well as the cubature scheme (which runs single-threaded using SciPy). We implemented custom, parallel assembly for computing elastic Hessians and gradients using Warp's Tile API to efficiently compute pairwise handle interactions in each element via our reduced cubature rules. All calculations are carried out in FP32 to avoid the steep performance penalty incurred by FP64 on consumer GPU hardware. All tetrahedral meshes were created using TetWild~\citep{Hu:2018:TMW:3197517.3201353}

All experiments were carried out on an NVIDIA DGX Spark mini-PC which features a 20-core (10 high-performance, 10 efficiency cores) ARM CPU on chip with a Blackwell GPU with $6,144$ shader cores. The CPU and GPU share 128 GB of unified LPDDR5x memory (273 GB/s memory bandwidth). The Spark runs DGX OS which is a custom Linux build based on Ubuntu 24.04. All simulation rendering was performed using Omniverse USD Composer~\citep{omniverse_kit_app_template} while the real-time bridge demo uses ovrtx~\citep{ovrtx_sdk} for rendering. 
 
\begin{figure*}[ht]
\includegraphics[width=\textwidth]{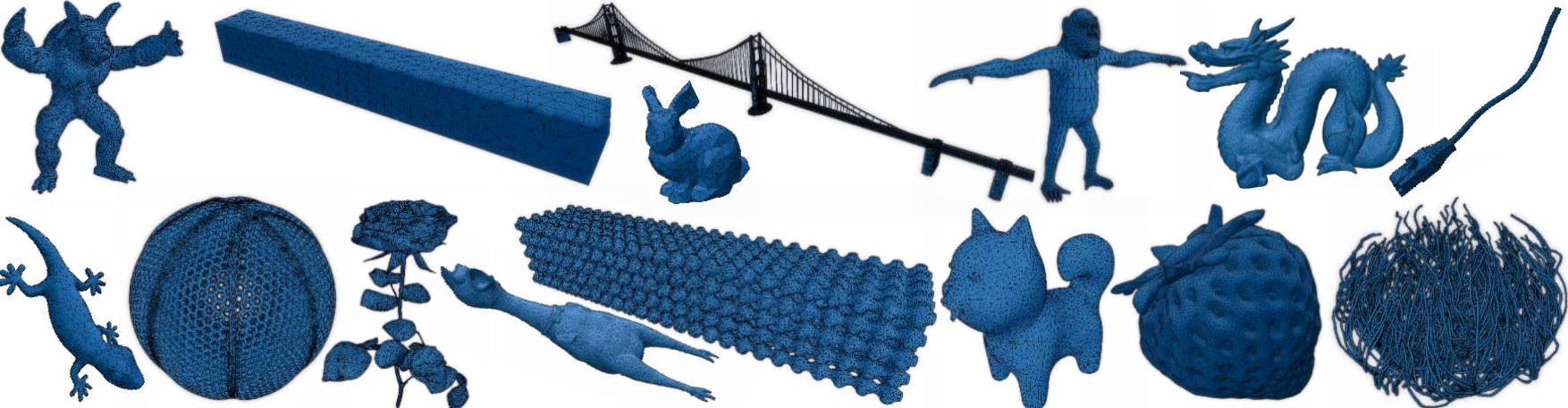}
\caption{The 15 tetrahedral meshes used for testing our decimation scheme ranging in complexity from $2000$ Vertices, $6000$ Tetrahedra (Bunny) to $200,000$ Vertices, $796,000$ Tetrahedra (Basketball) }
\Description{Gallery of example meshes.}
\label{fig:gallery}
\end{figure*}

In all cases we set $\hsupp = 6$, $H_{\min}= 7$ and the initial weight smoothness regularizer $\lambda = 0.01$. We give individual parameters for decimation error tolerance and number of modes to fit in \autoref{tab:decimation_stats_updated_2}. For all examples we use a cubature relative error of $10^{-3}$ and solve all linear systems using block-Jacobi Preconditioned  Conjugate Gradient (PCG) with an absolute tolerance of $10^{-6}$. All simulations are run with a timestep of $0.01$ seconds. We used 15 different meshes ranging in size and geometric complexity for testing our method~(\autoref{fig:gallery},~\autoref{tab:mesh_stats}).

\begin{table}[htbp]
  \centering
  \caption{\textbf{Mesh statistics.} Vertex and tetrahedra counts for various models. Figure~\ref{fig:gallery} shows images of each geometry}
  \label{tab:mesh_stats}. 
  
  % \rowcolors{start-row}{odd-row-color}{even-row-color}
  % Starting at row 2 so the header remains white
  \rowcolors{2}{white}{lightcornflower}
  
  \begin{tabular}{lrrr}
    \toprule
    \textbf{Mesh Name} & \textbf{Vertices ($|\vertices|$)} & \textbf{Tetrahedra ($|\tets|$)}  \\
    \midrule
    Armadillo    & $17,374$   & $68,089$ \\
    Beam & $7,000$   & $32,805$   \\
    Bridge       & $213,830$   & $725,505$  \\
    Chimpanzee         & $5,513$ & $20,111$  \\
    Coarse Bunny & $699$ & $2274$ \\
    Dragon & $41,796$ & $170,524$ \\
    Cable & $12,950$ & $52,513$  \\
    Gecko & $5,011$ & $14,250$  \\
    Basketball & $201,942$ & $796,623$ \\
    Rose & $50,296$ & $179,967$ \\
    Chicken & $10,663$ & $42,683$ \\
    Schwartz Beam & $167,786$ & $654,352$ \\
    Shiba & $9,101$ & $37,571$  \\
    Strawberry & $62,803$ & $276,297$ \\
    Tumbleweed & $99,751$ & $220,814$ \\
    \bottomrule
  \end{tabular}
\end{table}

\subsection{Decimation Evaluation}
\begin{figure*}
\includegraphics[width=\textwidth, trim={0 2mm 0 5mm}, clip]{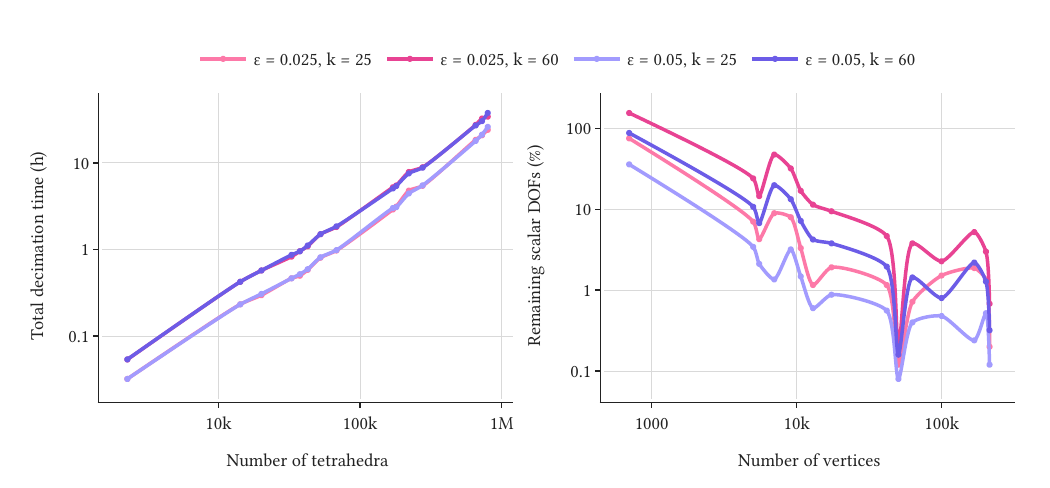}
\caption{Left: Total decimation time, in hours, as a function of mesh tetrahedra. Right: Percentage of vertex scalar degrees-of-freedom after decimation, as a function of mesh vertices. Both are log-log plots.}
\Description{Decimation time DOF reduction plots.}
\label{fig:decimate_stats}
\end{figure*}

The runtime of our algorithm grows proportionally with the number of tetrahedra in the input mesh~(\autoref{fig:decimate_stats}(left)), which is to be expected given the {\em one handle at a time}
% avoid scare quotes?
decimation strategy. Decreasing the error tolerance has virtually no effect on algorithm runtime. However, increasing the number of target displacement fields incurs a small performance penalty, likely because the current implementation loops over the modes when computing the fitting error and associated gradients and Hessians. 

\begin{figure}
\includegraphics[width=\columnwidth, trim={0 7mm 0 5mm}, clip]{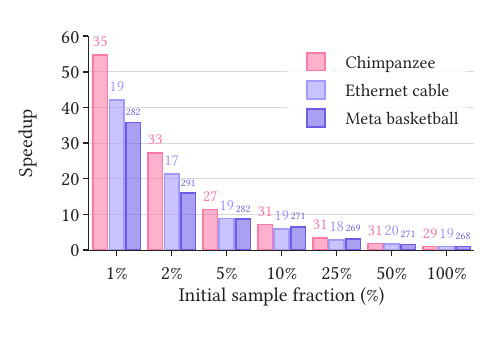}
\caption{Warmstarting our method with a subset of input tetrahedra reduces decimation time while still satisfying specified error thresholds and maintaining output handle counts (shown above data points).}
\Description{Chart showing almost 60x speedup with 1 percent initial sample fraction.}
\label{fig:warmstart}
\end{figure}

Because our method performs sequential handle removal it is readily amenable to warm starting with a subset of input tetrahedra~(\autoref{fig:warmstart}). We tested warms starts on three of our input meshes and observe an up to $54\times$ reduction in decimation times while maintaining the ability to reach user specified error tolerance ($\errgth = 0.05$) with similar final handle counts. Our input meshes tend to be high resolution in order to capture geometric detail, therefore even aggressive sub-sampling does not readily affect the spatial resolution of the initial handle arrangements, allowing the decimation scheme to remain effective. We observe sublinear scaling in speedup due to the increase in computation time for the initial weights as that linear system becomes denser for the larger handle neighborhoods in the subsampled input. For the remainder of the paper we demonstrate results using handles computed from the full input meshes as to provide rigorous analysis of the decimation scheme itself, however, these results point to warm-starting as an effective method of acceleration for production deployment of the method. Warmstarting Chimpanzee with $0.5\%$ of input tetrahedra (approximately $100$) leads to a discretization that already exceeds $\errgth$ at initialization, meaning decimation is required to get our low handle counts.

\autoref{fig:decimate_stats} (right) shows that, in general, our algorithm is more effective at reducing scalar DOF counts as input meshes become more complex. For low error threshold and high number of target fields, our method can actually increase the number of scalar DOFs for coarse meshes with small numbers of vertices. However, the general trend is that of significant reductions in DOF counts especially for moderate error counts and for complex meshes. For $\errgth = 0.05$, all meshes, except Coarse Bunny, see reductions of roughly $90\%$ or more in terms of scalar DOF counts. 

% --- Add these to your preamble if not already there ---
% \usepackage{booktabs}
% \usepackage[table]{xcolor} % Or \documentclass[acmtog, table]{acmart}

In \autoref{tab:brandt_comparison} we show an error comparison~(\autoref{sec:error}), under equal basis memory to \citet{brandt2017compressed}. In all cases our method produces lower error solutions than previous work, exceeding an order-of-magnitude reduction for complex meshes. While our method provides fixed handle influence per-vertex by construction, \citet{10.1145/3197517.3201387} relies on a geometric distance threshold which densifies the basis. Further, bases produced by prior work do not exhibit monotonic memory footprint reduction with respect to handle count. We use grid search followed by golden ratio line search to find appropriate handle counts in this case. In \autoref{sec:libarbic} we perform the same experiment with respect to ~\citet{li2019multi} which shows both worse maximum error reduction and average error reduction ($20.28\times$ and $8.34\times$ respectively). Our effective use of memory makes our algorithm particularly well-suited for hardware in which compute is "cheap" and memory is "expensive" such as modern GPUs.

\begin{table*}[htbp]
  \centering
  \caption{
  \textbf{Equal-Memory Subspace Accuracy Comparison.}
  Quantitative comparison between Brandt et al.~\cite{10.1145/3197517.3201387} and our modal handle decimation under equal basis memory allocation, sorted by error reduction factor. All tests evaluate $k=25$ deformation modes under corotational elasticity with a target tolerance $\text{tol}=0.05$. While Brandt et al. offer near-instantaneous offline construction, our method yields up to an order-of-magnitude higher accuracy for the same memory footprint.}
  \label{tab:brandt_comparison}
  
  % Starting at row 2 so the header remains white
  \rowcolors{2}{white}{lightcornflower}

  \setlength{\tabcolsep}{9.5pt} % PGK: eat up that white space!
  \begin{tabular}{l c c c c c}
    \toprule
    \textbf{Asset} & \textbf{Error Reduction} & \textbf{Memory (MB)} & \textbf{Handles ($h$)} & \textbf{Subspace Error ($\mathcal{E}$)} & \textbf{Subspace Error ($\mathcal{E}$)} \\
    & $\left(\mathcal{E}_{\text{Brandt}} / \mathcal{E}_{\text{Ours}}\right)$ & & \textbf{Brandt / Ours} & \textbf{Brandt et al. [2018]} & \textbf{Ours} \\
    \midrule
    Tumbleweed      & \textbf{15.60$\times$} & 82.2  & 6 / 123   & 0.3784 & 0.0243 \\
    Meta Basketball & \textbf{15.12$\times$} & 166.4 & 8 / 268   & 0.2482 & 0.0164 \\
    Bridge          & \textbf{3.67$\times$}  & 176.2 & 17 / 58   & 0.0730 & 0.0199 \\
    Schwartz Beam   & \textbf{2.83$\times$}  & 138.3 & 10 / 95   & 0.0908 & 0.0321 \\
    Beam            & \textbf{1.84$\times$}  & 5.8   & 13 / 59   & 0.0372 & 0.0202 \\
    Chimpanzee      & \textbf{1.64$\times$}  & 4.5   & 29 / 29   & 0.0334 & 0.0204 \\
    Ethernet Cable  & \textbf{1.56$\times$}  & 10.7  & 13 / 19   & 0.0330 & 0.0211 \\
    \midrule
    \textbf{Average}& \textbf{6.04$\times$}  & 83.4  & 13.7 / 93.0 & 0.1277 & 0.0220 \\
    \bottomrule
  \end{tabular}
\end{table*}

\subsection{Cubature Evaluation}
\begin{figure}
\includegraphics[width=\columnwidth, trim={0 2mm 0 5mm}, clip]{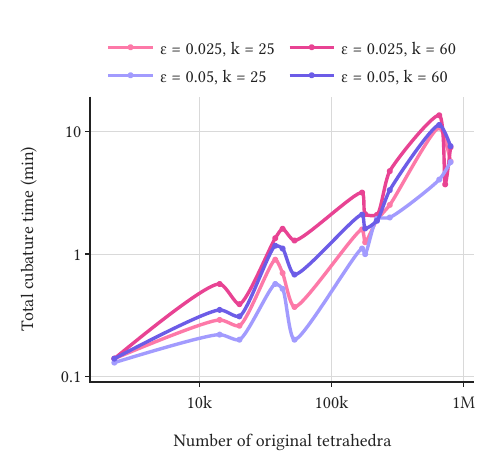}
\caption{Time, in minutes to compute optimal cubature rules for all example meshes.
}
\Description{Cubature compute time as a function of number of tetrahedra.}
\label{fig:cube_perf}
\end{figure}

Our reduced cubature also exhibits runtime that grows proportionally with the number of input tetrahedra~(\autoref{fig:cube_perf}). The relationship is noisier due to the iterative nature of the algorithm and the fact that thin shapes, like the rose~(\autoref{tab:cubature_stats}), often require fewer cubature points than bulkier geometries, even when they possess more input tetrahedra (required to capture thin features). As with decimation time, we see a mild to moderate increase in times as error tolerance is decreased and number of target vector fields increased. Overall our cubature algorithm is reasonably quick, taking a maximum of $13.60$ minutes on our large Schwartz beam example ($654,352$ tetrahedra). Compactly supported bases necessitate more cubature samples than those with global support since each cubature tetrahedron only samples a subset of the full basis; however, assembly performance is still fast since the disjoint nature of the basis functions allows for fast, parallel assembly~(\autoref{fig:sim_timings}). In our examples the time taken for assembly of the global system matrix is always significantly less than $1$ millisecond ($\errgth=0.05$ and $\nummodes=25$), regardless of input system size.

The relationship between the remaining percentage of tetrahedra comprising the optimal cubature rule and the input tetrahedra count is more murky. As meshes become larger, our method yields relatively smaller cubature rules; however, some geometries, like the beam, require significantly more cubature tets to accurately integrate the weight functions and forces generated by random sampling the frame DOFs. As expected, lowering error and increasing target vector field count causes the number of cubature tetrahedra to increase as the cubature rule must accurately integrate more functions due to the additional DOFs produced during decimation, as shown in \autoref{tab:decimation_stats_updated_2}. In all cases, we achieve a reduction in tetrahedra required for cubature with even our most extreme cases, on coarse meshes requiring roughly $80\%$ of the input tetrahedra count to achieve accurate integration of elastic gradients and Hessians. 

\begin{figure}
\includegraphics[width=\columnwidth, trim={0 2mm 0 5mm}, clip]{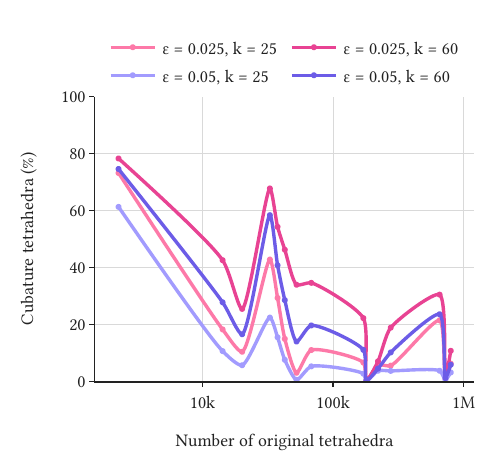}
\label{fig:cube_output}
\caption{Percentage of tetrahedra used for optimal cubature as a function of input mesh size in tetrahedra.
}
\Description{Plot showing minimal percentage of cubature tetrahedra for big meshes.}
\end{figure}

\subsection{Simulation Evaluation}
\begin{figure}
\includegraphics[width=\columnwidth, trim={0 2mm 0 5mm}, clip]{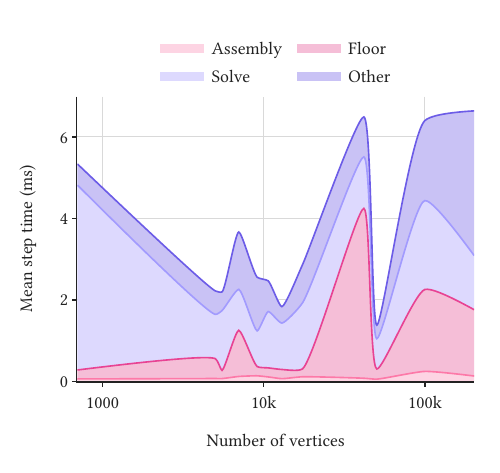}
\caption{Mean per-timestep timings, in milliseconds for drop tests using error at 0.05 and 25 target modal vector fields.}
\Description{Plot showing simulation run times.}
\label{fig:sim_timings}
\end{figure}

We evaluate runtime simulation performance of our method on all meshes, with error $0.05$ and $25$ target displacement fields which we find sufficient to achieve good visual quality~(see figures~\ref{fig:armadillo} and onward as well as the accompanying supplemental video). In all cases, even those where the number of reduced cubature tetrahedra can spike, performance is excellent, with step times rarely exceeding $5$ milliseconds~(\autoref{fig:sim_timings}). In all cases either the PCG solver or floor contact is the bottleneck. We do not implement broadphase collision detection and so meshes with many surface vertices have relatively worse floor contact handling runtime. Because penalty springs are processed for contacting vertices only, meshes such as the beam, where a large percentage of surface vertices contact the floor at once, incur a performance penalty. Performance of our solver is partly influenced by the tight tolerance we solve to, however even with these addressable bottlenecks, our runtime performance is fast. 

\begin{figure*}[ht]
\includegraphics[width=\textwidth]{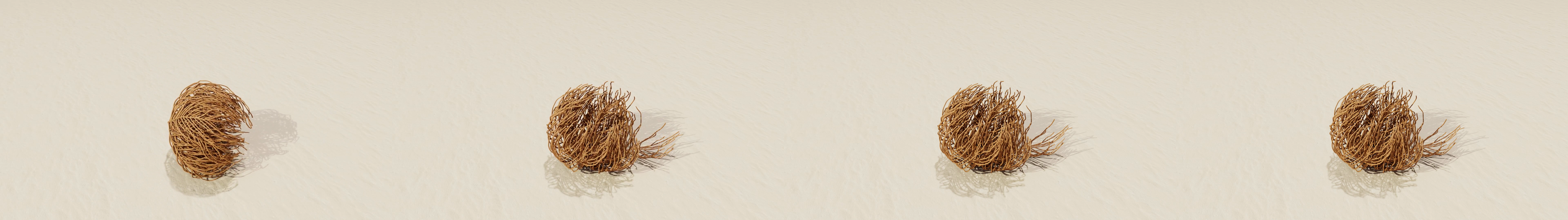}
\caption{Real-time simulation of the complex Tumbleweed geometry falling onto the desert floor.}
\Description{Four images from simulation of the Tumbleweed mesh.}
\label{fig:tumbleweed}
\end{figure*}

There is a performance penalty, attributed to the higher scalar DOF counts, for decreasing decimation tolerance or increasing the number of target vector fields~(\autoref{tab:performance_stats_updated}). With $25$ modes, decreasing the decimation error tolerance to $0.025$ incurs a $1.68\times$ increase in mean frame time ($2.80\times$ at $60$ modes). Similarly, increasing the number of target vector fields from $25$ to $60$ incurs a $1.01\times$ mean performance hit at $0.05$ decimation error and $1.69\times$ at $0.025$ decimation error, respectively. Given that increasing the number of target vector fields is not particularly punitive, performance-wise, it is a practically useful adjustment to capture even finer details in deformation simulation. The Shiba example (\autoref{fig:shiba}) uses $60$ target vector fields to capture the bending of the dog's tongue as it comes in contact with the ground plane, a small detail difficult to capture with standard reduced-order approaches. Even without resorting to higher target vector field counts, our decimated handles can capture compelling deformations of complicated aggregate thin structures like this tumbleweed~(\autoref{fig:tumbleweed}).

\begin{figure}[t]
\includegraphics[width=\columnwidth]{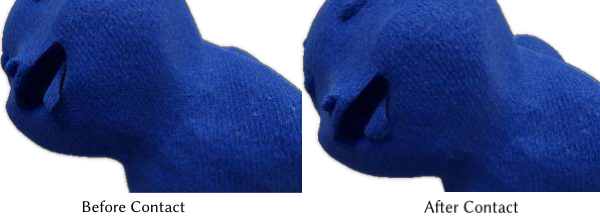}
\caption{Simulation of the Shiba mesh (error $0.05$, modes $60$). Left: The undeformed tongue of the Shiba mesh prior to contact. Right: Our handles and weights capture the bending of the tongue on contact with the ground (seen from below the ground plane)}
\Description{Close up renders of Shiba before and after contact.}
\end{figure}

\subsection{Real-time Applications}
\begin{figure*}
\includegraphics[width=\textwidth]{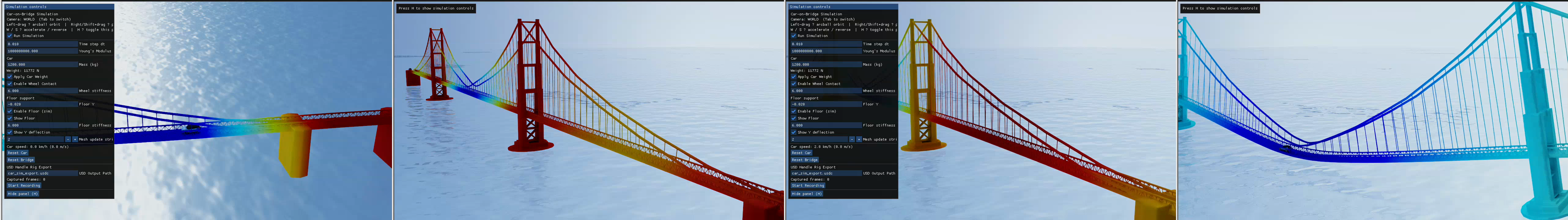}
\caption{Screen captures from an interactive application in which a user drives a car across a fully simulated Golden Gate Bridge. Colors indicate vertical displacement of the bridge. In the last frame, the mass of car has been increased to $5000$kg to amplify deflection. Note that the bridge is not pinned to the ground, we use penalty spring contact.}
\Description{Four images of a bridge model undergoing deformation.}
\label{fig:interactive}
\end{figure*}

\begin{figure*}
\includegraphics[width=\textwidth]{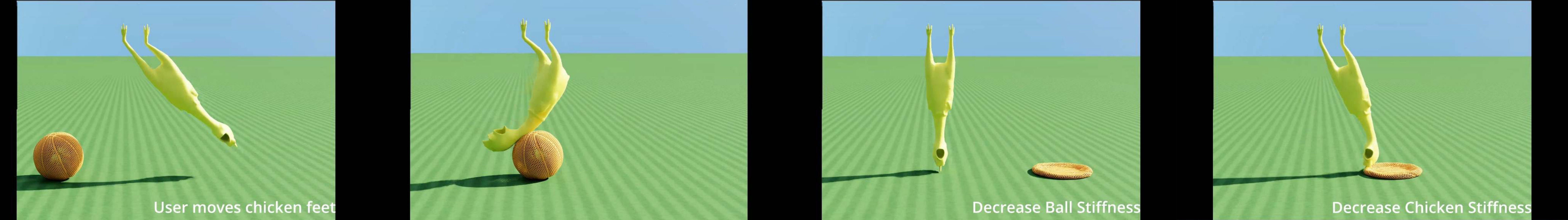}
\caption{Boccer: Screen capture from an interactive application in which a user manipulates the feet of an inverted rubber chicken in an attempt to push a meta-material basketball along the ground. The stiffness of each object can be adjusted in real-time to achieve various effects.}
\Description{Four images of a rubber chicken playing with a metamaterial ball.}
\label{fig:boccer}
\end{figure*}

The performance offered by our approach makes the generated reduced-order assets well-suited to real-time, interactive applications. \autoref{fig:interactive} shows an interactive demo wherein a user can accelerate or decelerate a car along the Golden Gate Bridge. The entire bridge is simulated using our handle-based approach. The car is modeled as a particle with mass. Colors show the deflection of the bridge in the y-direction (blue negative, red postive). The bridge stiffness is only $10^9$ Pascals (Pa) to exaggerate the deformations for visual effect. Half way through the interaction, we increase the mass of the car to $5000$ kilograms (kg) to further distort the bridge. Compact support in our basis functions means that our handle-based reduction does not bake boundary conditions into the reduced basis, rather the contact between the bridge and the water textured ground plane is handled with penalty springs (again in real-time). This entire application, including rendering, runs at $30$ fps on the aforementioned DGX Spark system. See our supplemental video for a full screen capture. We further demonstrate the applicability of our handles to the interactive regime by creating a small game we call Boccer (a portmanteau of the chicken sound, ``Ba-Kaw'', and soccer)~(\autoref{fig:boccer}). Here a user manipulates the fixed feet of the rubber chicken in an attempt to push the meta-material basketball along the ground. The stiffness of both objects is adjusted in real-time to achieve different effects. Both objects are simulated using our handles and contact is handled via penalty springs.

%% file: sections/conclusion.tex
%We have presented a novel decimation approach to computing handle-based discretization with sparse, compactly supported weights that is far more memory efficient then previous works~(\autoref{tab:brandt_comparison}).
% PGK: adjusted for overfull hbox :(
We present a novel decimation approach for handle-based discretization. By using sparse, compactly supported weights, our method is significantly more memory-efficient than prior work~(\autoref{tab:brandt_comparison}).
We have demonstrated the ability of our algorithm to compute such discretizations on a variety of large, complicated geometries, the largest of which exceeds three-quarters of a million tetrahedra.
We have also shown that the resulting compactly supported weights can be exploited to yield elastodynamics simulations which run far in excess of real-time rates, with most of our examples exceeding the 100 frames-per-second threshold.

\subsection{Limitations}
While highly optimal for deployment in game engines and learning loops, our strict error-guided decimation is computationally heavy, taking up to 36 hours in the worst case. As our method can be warm-started with an initial, smaller set of handles, any upfront sampling technique can be applied to dramatically improve preprocessing time~(e.g., 30 to 60 times faster with minimal impact on the final model, \autoref{fig:warmstart}). 
Further, for the applications we are targeting, such as repeated interactive simulation, such as robotics or games where a single object can be resimulated countless times. In these scenarios, preprocessing time arguably becomes an afterthought. 
However, further performance improvements are an important direction of future work.
The handle-based representation produced by decimation is not necessarily optimal for geometries that are well-represented by coarse meshes.
Coarse Bunny achieves at most a $64\%$ reduction in DOFs and, for low error and high modes, actually has more DOFs than just simulating using vertices (a similar trend is observed for the optimal cubature rules)~(\autoref{tab:decimation_stats_updated_2},~\autoref{tab:cubature_stats}).
We note this is a rare occurrence when focusing on complex geometries (only $1$ of our $60$ decimated examples).
Beyond that, unlike some recent methods, our approach is not data-free~\citep{Simplicits2024} and requires computation of a target set of displacement vector fields.
Previous work has shown that matching linear modes is a reasonable assumption for computing coarse discretizations but relaxing this assumption may make the method easier to apply to certain learning-based simulation paradigms.
Finally, our approach maintains a fixed support size for each vertex.
Allowing this support to shrink dynamically would enable more compact representations of rigid parts where a single handle can represent the motion.

\subsection{Future Work}
Beyond performance improvements and a data-free formulation of decimation, we note that the lion's share of our runtime per timestep is spent in the block-Jacobi preconditioned CG solver.
While we could reduce the solver tolerance, finding alternative faster-converging approaches would be beneficial.
It is also important to move away from requiring a well-formed input tetrahedral mesh, because many modern 3D shape representations, such as Gaussian splats, do not readily admit tetrahedral meshes.
Applying our technique to these types of data likely requires a reformulation of the decimation scheme to act on an underlying particle distribution or to mimic modern Gaussian splat fitting schemes, which rely on Monte Carlo optimization~\citep{kheradmand20243d}, rather than strict hill descent.
We hope that the flexibility, expressivity, and performance we have illustrated here will encourage the community to further explore both decimation-type and sampling-type approaches to simultaneous handle and weight computation.
To encourage future work, we intend to release all code, data, computed handles and cubature rules.

%% file: sections/appendices.tex
\section{Reduced-Order Equations of Motion}
\label{sec:rom}

The standard equations governing the dynamics of a discretized elastic solid are typically written as 
\begin{equation}
M\ddot{\mathbf{q}} = \mathbf{f}(\mathbf{q}).
\label{eq:eom}
\end{equation} where $\mathbf{q}$ is the stacked vector of system degrees-of-freedom, $M$ is the mass-matrix and $\mathbf{f}$ is the force which can include forces due to elasticity and body forces. Here we ignore forces proportional to velocity such as viscosity. 

Classical reduced-order models operate within a linear, lower-dimension reduced space which can be projected back via
\begin{equation}
\mathbf{q} = \Phi\mathbf{r},
\label{eq:reduced_space}
\end{equation} where $\Phi\in\mathbb{R}^{n\times r}$ and $\mathbf{r}\in\mathbb{R}^r$ where $r \ll n$.
In our case, $\Phi$ consists of the lowest-frequency eigenvectors of the generalized eigenvalue problem obtained by linearizing \autoref{eq:eom} at rest.
The Galerkin projection of \autoref{eq:eom} into our reduced space yields the following equation 
\begin{equation}
\Phi^TM\Phi\ddot{\mathbf{r}} = \Phi^T\mathbf{f}(\Phi\mathbf{r}).
% PGK: added the missing \Phi^T projection of forces
\label{eq:reom}
\end{equation} 
% PGK: suppressed the false statement below, and added revised statement after 16.
%In our case $\Phi$ is a sparse matrix created using our compactly-supported skinning handle weights. 
    
\section{Decimation Preserves Partition of Unity}
% PGK: no dashes when a compound modifier (adjective) but not when used as a noun (as below)
\label{sec:pou}
For any vertex $v$, prior to removing a supporting handle $\handle^*$ we have $\sum_{h\in\supp{v}} \wfun_{vh} = 1$.
After removing $\handle^*$,
% PGK!!
%deformation 
we compute a new support, $\newsupp{v}$ where all but one handle is identical to those in $\supp{v}$
%, $\handle^*$, % This was tilde before, denoting the NEW handle... 
% Is the stuff likewise below broken?
%\paul{I might have broken this a while back... do we need to reference the new handle dumped into the list by Yen's alg?  I don't think so, in which case delete this comment!} is identical to those in $\supp{v}$. 
Due to compact support $\wfun_{v\handle^*} = 0$. Therefore,
\begin{equation}
\sum_{h\in\newsupp{v}} \wfun_{vh} = \sum_{h\in(\newsupp{v}\backslash\handle^*)}\wfun_{vh} = 1 - \wfun_{v\handle^*}, 
\end{equation} 
where $\wfun_{v\handle^*}$ is $\handle^*$'s pre-removal weight at vertex $\vertex$. 

Our algorithm redistributes $\wfun_{v\handle^*}$ to handles in $\newsupp{v}$ and so the partition of unity at $v$ becomes
\begin{align}
  &\phantom{=} \ \  \sum_{h\in\newsupp{v}} \left(\wfun_{vh} + \alpha_h\wfun_{v\handle^*}\right) \\
  &= 1 - \wfun_{v\handle^*} + \wfun_{v\handle^*}\sum_{h\in\newsupp{v}}\alpha_h \\
  &= 1 - \left( 1-\sum_{h\in\newsupp{v}}\alpha_h \right) \wfun_{v\handle^*}
\end{align}
where $\alpha_h$ is the redistribution weight.

Partition of unity requires $1 -(1-\sum_{h\in\newsupp{v}}\alpha_h)\wfun_{v\handle^*} = 1$ which implies $(1-\sum_{h\in\newsupp{v}}\alpha_h)\wfun_{v\handle^*} = 0$. For arbitrary $\alpha_h$ and $\wfun_{v\handle^*}$ this requires $\sum_{h\in\newsupp{v}}\alpha_h = 1$ which is exactly the constraint we apply in \autoref{sec:update}.

% Placed here for page flow..
\input{sections/Table_Li_Barbic}

%%%%%%%%%%%%%%%%%%%%%%%%%%%%%%%%%%%%%%%%%%%%%%%%%%%%%%%%%%%%
\section{Derivation of Handle Formulation} 
\label{sec:disp}

We rephrase the positional handle equation~(\autoref{eq-skinning}) into a compact displacement form, under the assumption of partition of unity.
Substituting from ~\autoref{eq-skinning} into $\mathbf{u}(\refx) = \defx(\refx) - \refx$, we have
\begin{equation}
\mathbf{u}(\refx) = \sum_{\sphandle\in\handleset}\wfun_\sphandle(\refx)\Taffine_{\sphandle}{}\begin{bmatrix}(\refx-\hori_\sphandle)\\1\end{bmatrix} - \refx.
\end{equation}
Rewriting $\refx$ into a compatible handle form, we obtain
\begin{equation}
\mathbf{u}(\refx) = \sum_{\sphandle\in\handleset}\wfun_\sphandle(\refx)\Taffine_{\sphandle}{}\begin{bmatrix}(\refx-\hori_\sphandle)\\1\end{bmatrix} - \begin{bmatrix}I & 0\end{bmatrix}\begin{bmatrix}\refx\\ 1 \end{bmatrix},
\end{equation} where $I$ is a $3\times 3$ Identity matrix. Assuming partition of unity, 
\begin{equation}
\nonumber
\mathbf{u}(\refx) = \sum_{\sphandle\in\handleset}\wfun_\sphandle(\refx)\Taffine_{\sphandle}{}\begin{bmatrix}(\refx-\hori_\sphandle)\\1\end{bmatrix} - \sum_{\sphandle\in\handleset}\wfun_\sphandle(\refx)\begin{bmatrix}I & \mathbf{C}_h\end{bmatrix}\begin{bmatrix}\refx - \mathbf{C}_h\\ 1 \end{bmatrix},
\end{equation} and then collecting terms yields
\begin{equation}
\label{eq-skinning-displ}
\mathbf{u}(\refx) = \sum_{\sphandle\in\handleset}\wfun_\sphandle(\refx)\Uaffine_{\sphandle}{}\begin{bmatrix}(\refx-\hori_\sphandle)\\1\end{bmatrix},
\end{equation} where $\Uaffine_h = \begin{bmatrix}\mathbf{A}_h-I && \mathbf{p}_h - \mathbf{C}_h\end{bmatrix}$, $\mathbf{A}_h$ is the $3\times3$ linear component of $\Taffine_h$ and $\mathbf{p}_h$ is the $3\times1$ translation component, ie $\Uaffine$ stores the displacement away from the reference handle state.

\section{Comparison Error} 
\label{sec:error}

We compare the accuracy of our method to previous work  using normalized, mass-weighted error, computed between all target displacement fields and the best least-squares, mass-weighted reconstruction of those fields in the basis ($B$) of the given method.
\begin{align}
E&=\sqrt{\sum_{v=0}^{|\vertices|-1}\Biggl(\frac{1}{k}\sum_{m=0}^{k-1}\frac{\sqrt{\rho_v\|u_v^m-(B\mathbf{z}^{m})_v\|_2^2}} {\sqrt{\sum_{v'=0}^{|\vertices|-1}\rho_{v'}\|u_{v'}^{(k)}\|_2^2}}\Biggr)^{\!2}}, \\
\mathbf{z}^{m}&=\arg\min_z\|\mathbf{u}^m-B\mathbf{z}\|_M,
\end{align}

% PGK: I'm trying to wrangle these end pages... Sorry for the damage, but perhaps there is a nice solution??  

%\FloatBarrier
% These two appendices contain only full-width floats. In two-column mode a
% figure*/table* can never land on the page it is written on, so each heading
% was stranded on an otherwise blank page. Outside two-column mode the starred
% floats become ordinary \textwidth floats and pack in beneath their heading.

%\onecolumn

%\FloatBarrier
\section{Comparison to Li and Barbic 2019}
\label{sec:libarbic}

We compare modal fitting error~(\autoref{sec:error}) to the hierarchical, compactly-supported subspaces of \citet{li2019multi} (our own implementation since no author implementation is publicly available) under equal memory conditions. To find the optimal hierarchy setup, we sweep both the level 0 basis size and the hierarchy depth to find the setup that best minimizes modal fitting error under the prescribed memory usage.

\section{Performance Tables}

\autoref{tab:decimation_stats_updated_2} shows decimation statistics for different models with different thresholds.  \autoref{tab:cubature_stats} shows cubature statistics for different models and varying error thresholds. \autoref{tab:performance_stats_updated} shows simulation timings for different models and different numbers of modes.

\section{Simulation Frames}

\autoref{fig:armadillo} through \autoref{fig:strawberry} show example images from a variety of real-time simulations that demonstrate our method, while \autoref{fig:chicken1} and \autoref{fig:chicken2} show example simulations of varying material stiffness.  See also the supplementary video.

\input{sections/Appendix_Perf_Tables}

\input{sections/Appendix_Screengrabs}

%% file: sections/Table_Li_Barbic.tex
\begin{table*}[t]
  \centering
  \caption{\textbf{Equal-Memory Subspace Accuracy Comparison (Li \& Barbič).} Quantitative comparison between Li and Barbič~\cite{li2019multi} and our modal handle decimation under equal basis memory allocation, sorted by error reduction factor. All tests evaluate $k=25$ deformation modes under corotational elasticity with a target tolerance $\text{tol}=0.05$.}
  \label{tab:libarbic_comparison}
  
  \rowcolors{3}{white}{lightcornflower}

\setlength{\tabcolsep}{9pt} % PGK: eat up that white space!

  \begin{tabular}{l c c c c c}
    \toprule
    \textbf{Asset} & \textbf{Error Reduction} & \textbf{Memory (MB)} & \textbf{Handles ($h$)} & \textbf{Subspace Error ($\mathcal{E}$)} & \textbf{Subspace Error ($\mathcal{E}$)} \\
    & $\left(\mathcal{E}_{\text{Li\&Barbič}} / \mathcal{E}_{\text{Ours}}\right)$ & & \textbf{Li \& Barbič / Ours} & \textbf{Li \& Barbič} & \textbf{Ours} \\
    \midrule
    Tumbleweed      & \textbf{20.28$\times$} & 82.2  & 62 / 123  & 0.4917 & 0.0243 \\
    Meta Basketball & \textbf{10.52$\times$} & 166.4 & 69 / 268  & 0.1727 & 0.0164 \\
    Bridge          & \textbf{9.07$\times$}  & 176.2 & 161 / 58  & 0.1803 & 0.0199 \\
    Schwartz Beam   & \textbf{7.69$\times$}  & 138.3 & 25 / 95   & 0.2466 & 0.0321 \\
    Ethernet Cable  & \textbf{4.46$\times$}  & 10.7  & 82 / 19   & 0.0942 & 0.0211 \\
    Beam            & \textbf{3.79$\times$}  & 5.8   & 80 / 59   & 0.0765 & 0.0202 \\
    Chimpanzee      & \textbf{2.59$\times$}  & 4.5   & 207 / 29  & 0.0527 & 0.0204 \\
    \midrule
    \rowcolor{white}
    \textbf{Average}& \textbf{8.34$\times$}  & 83.4  & 98.0 / 93.0 & 0.1878 & 0.0220 \\
    \bottomrule
  \end{tabular}
\end{table*}

%% file: sections/Appendix_Perf_Tables.tex
%\FloatBarrier

% using the long tables, stretch them a tiny bit to better flow on the pages
\renewcommand{\arraystretch}{1.05}

\input{sections/Table_Decimation_Stats_LongTable}

\input{sections/Table_Cubature_Stats_Longtable}

\input{sections/Table_Perf_LongTable}

%% file: sections/Table_Decimation_Stats_LongTable.tex
\clearpage
\onecolumn

% Start alternating row colors
\begingroup
  \setlength{\tabcolsep}{5.7pt}

\rowcolors{3}{white}{lightcornflower}

\begin{longtable}{lrrrrrrrr}
    % --- 1. CAPTION & HEADER FOR FIRST PAGE ---
    \caption{
    \textbf{Decimation statistics.}
     Thresholds, handle counts, percentages, and execution times for all experiments.} 
    \label{tab:decimation_stats_updated_2} \\
    \toprule \rowcolor{white}
    & & & & & & \multicolumn{1}{c}{\textbf{Initial}} & \multicolumn{1}{c}{\textbf{Init}} & \\
    \rowcolor{white}
    \multirow{-2}{*}{\textbf{Example}} & \multirow{-2}{*}{\textbf{Threshold}} & \multirow{-2}{*}{\textbf{\# Modes}} & \multirow{-2}{*}{\textbf{\# Handles}} & \multirow{-2}{*}{\textbf{Percentage}} & \multirow{-2}{*}{\textbf{Modes (m)}} & \textbf{Weights (m)} & \textbf{Queue (m)} & \multirow{-2}{*}{\textbf{Decimate (h)}} \\
    \midrule
    \endfirsthead

    % --- 2. HEADER FOR SUBSEQUENT PAGES ---
    \rowcolor{white}\caption*{(continued)} \\
    \toprule\rowcolor{white}
    & & & & & & \multicolumn{1}{c}{\textbf{Initial}} & \multicolumn{1}{c}{\textbf{Init}} & \\
    \multirow{-2}{*}{\textbf{Example}} & \multirow{-2}{*}{\textbf{Threshold}} & \multirow{-2}{*}{\textbf{\# Modes}} & \multirow{-2}{*}{\textbf{\# Handles}} & \multirow{-2}{*}{\textbf{Percentage}} & \multirow{-2}{*}{\textbf{Modes (m)}} & \textbf{Weights (m)} & \textbf{Queue (m)} & \multirow{-2}{*}{\textbf{Decimate (h)}} \\
    \midrule
    \endhead

    % --- 3. FOOTER FOR ALL PAGES EXCEPT LAST ---
    \midrule
    \rowcolor{white}
     \multicolumn{9}{r}{\small\itshape Continued on next page...} \\
    \endfoot

    % --- 4. FOOTER FOR LAST PAGE ---
    \bottomrule
    \endlastfoot

    % --- ALL TABLE DATA MERGED TOGETHER ---
    armadillo & 0.05 & 25 & 39 & 0.90\% & 0.10 & 0.34 & 2.90 & 0.93 \\
    armadillo & 0.05 & 60 & 165 & 3.80\% & 0.13 & 0.34 & 5.77 & 1.75 \\
    armadillo & 0.025 & 25 & 84 & 1.93\% & 0.10 & 0.34 & 2.89 & 0.92 \\
    armadillo & 0.025 & 60 & 412 & 9.49\% & 0.13 & 0.34 & 5.76 & 1.72 \\
    beam & 0.05 & 25 & 59 & 1.36\% & 0.04 & 0.11 & 1.39 & 0.44 \\
    beam & 0.05 & 60 & 348 & 19.89\% & 0.08 & 0.11 & 2.78 & 0.82 \\
    beam & 0.025 & 25 & 157 & 8.97\% & 0.04 & 0.11 & 1.40 & 0.44 \\
    beam & 0.025 & 60 & 834 & 47.66\% & 0.07 & 0.11 & 2.77 & 0.78 \\
    bridge & 0.05 & 25 & 58 & 0.11\% & 1.32 & 30.75 & 34.40 & 20.14 \\
    bridge & 0.05 & 60 & 173 & 0.32\% & 1.74 & 27.98 & 65.64 & 28.88 \\
    bridge & 0.025 & 25 & 115 & 0.22\% & 1.32 & 30.58 & 34.03 & 19.80 \\
    bridge & 0.025 & 60 & 361 & 0.68\% & 1.74 & 27.88 & 70.44 & 30.97 \\
    chimpanzee & 0.05 & 25 & 29 & 2.10\% & 0.03 & 0.07 & 0.96 & 0.29 \\
    chimpanzee & 0.05 & 60 & 93 & 6.75\% & 0.03 & 0.07 & 1.91 & 0.54 \\
    chimpanzee & 0.025 & 25 & 59 & 4.28\% & 0.02 & 0.07 & 0.94 & 0.28 \\
    chimpanzee & 0.025 & 60 & 201 & 14.58\% & 0.03 & 0.07 & 1.90 & 0.54 \\
    coarse\_bunny & 0.05 & 25 & 63 & 36.05\% & 0.00 & 0.01 & 0.11 & 0.03 \\
    coarse\_bunny & 0.05 & 60 & 154 & 88.13\% & 0.00 & 0.01 & 0.22 & 0.05 \\
    coarse\_bunny & 0.025 & 25 & 132 & 75.54\% & 0.00 & 0.01 & 0.11 & 0.03 \\
    coarse\_bunny & 0.025 & 60 & 272 & 155.65\% & 0.00 & 0.01 & 0.21 & 0.05 \\
    dragon & 0.05 & 25 & 59 & 0.56\% & 0.27 & 1.48 & 7.71 & 2.86 \\
    dragon & 0.05 & 60 & 205 & 1.96\% & 0.35 & 1.51 & 15.16 & 4.80 \\
    dragon & 0.025 & 25 & 121 & 1.16\% & 0.26 & 1.53 & 7.57 & 2.75 \\
    dragon & 0.025 & 60 & 491 & 4.70\% & 0.35 & 1.46 & 14.52 & 4.97 \\
    ethernet\_cable & 0.05 & 25 & 19 & 0.59\% & 0.07 & 0.23 & 2.40 & 0.77 \\
    ethernet\_cable & 0.05 & 60 & 137 & 4.23\% & 0.10 & 0.23 & 4.80 & 1.42 \\
    ethernet\_cable & 0.025 & 25 & 38 & 1.17\% & 0.07 & 0.23 & 2.41 & 0.76 \\
    ethernet\_cable & 0.025 & 60 & 371 & 11.46\% & 0.10 & 0.24 & 4.75 & 1.42 \\
    gecko & 0.05 & 25 & 43 & 3.43\% & 0.02 & 0.05 & 0.70 & 0.22 \\
    gecko & 0.05 & 60 & 135 & 10.78\% & 0.02 & 0.05 & 1.40 & 0.40 \\
    gecko & 0.025 & 25 & 88 & 7.02\% & 0.02 & 0.05 & 0.72 & 0.22 \\
    gecko & 0.025 & 60 & 302 & 24.11\% & 0.02 & 0.05 & 1.39 & 0.40 \\
    meta\_basketball & 0.05 & 25 & 268 & 0.53\% & 6.32 & 36.69 & 41.85 & 24.80 \\
    meta\_basketball & 0.05 & 60 & 640 & 1.27\% & 4.79 & 31.84 & 76.57 & 35.97 \\
    meta\_basketball & 0.025 & 25 & 691 & 1.37\% & 4.43 & 31.24 & 37.38 & 22.98 \\
    meta\_basketball & 0.025 & 60 & 1,519 & 3.01\% & 4.18 & 27.82 & 70.62 & 32.62 \\
    rose & 0.05 & 25 & 12 & 0.10\% & 0.39 & 1.87 & 7.87 & 2.92 \\
    rose & 0.05 & 60 & 21 & 0.17\% & 0.44 & 1.97 & 16.06 & 5.11 \\
    rose & 0.025 & 25 & 15 & 0.12\% & 0.37 & 1.95 & 8.07 & 2.96 \\
    rose & 0.025 & 60 & 26 & 0.21\% & 0.46 & 1.96 & 16.21 & 5.19 \\
    rubber\_chicken & 0.05 & 25 & 39 & 1.46\% & 0.05 & 0.17 & 1.82 & 0.56 \\
    rubber\_chicken & 0.05 & 60 & 192 & 7.20\% & 0.07 & 0.17 & 3.57 & 1.05 \\
    rubber\_chicken & 0.025 & 25 & 89 & 3.34\% & 0.05 & 0.17 & 1.77 & 0.55 \\
    rubber\_chicken & 0.025 & 60 & 452 & 16.96\% & 0.07 & 0.17 & 3.56 & 1.03 \\
    schwartz & 0.05 & 25 & 95 & 0.23\% & 7.45 & 18.87 & 28.46 & 17.17 \\
    schwartz & 0.05 & 60 & 926 & 2.21\% & 8.13 & 21.15 & 57.69 & 25.78 \\
    schwartz & 0.025 & 25 & 794 & 1.89\% & 7.28 & 21.14 & 29.60 & 17.65 \\
    schwartz & 0.025 & 60 & 2,199 & 5.24\% & 8.18 & 19.10 & 57.37 & 26.27 \\
    shiba & 0.05 & 25 & 73 & 3.21\% & 0.05 & 0.14 & 1.57 & 0.49 \\
    shiba & 0.05 & 60 & 304 & 13.36\% & 0.07 & 0.14 & 3.08 & 0.90 \\
    shiba & 0.025 & 25 & 182 & 8.00\% & 0.05 & 0.14 & 1.54 & 0.47 \\
    shiba & 0.025 & 60 & 730 & 32.08\% & 0.07 & 0.14 & 3.17 & 0.90 \\
    strawberry & 0.05 & 25 & 63 & 0.40\% & 2.11 & 3.32 & 11.97 & 5.25 \\
    strawberry & 0.05 & 60 & 224 & 1.43\% & 2.41 & 3.49 & 23.45 & 8.37 \\
    strawberry & 0.025 & 25 & 114 & 0.73\% & 2.10 & 3.28 & 11.57 & 5.20 \\
    strawberry & 0.025 & 60 & 595 & 3.79\% & 2.51 & 3.32 & 23.42 & 8.47 \\
    tumbleweed & 0.05 & 25 & 123 & 0.49\% & 0.27 & 4.52 & 11.07 & 4.18 \\
    tumbleweed & 0.05 & 60 & 202 & 0.81\% & 0.42 & 4.93 & 22.35 & 7.12 \\
    tumbleweed & 0.025 & 25 & 384 & 1.54\% & 0.28 & 4.97 & 11.84 & 4.51 \\
    tumbleweed & 0.025 & 60 & 566 & 2.27\% & 0.43 & 4.95 & 22.99 & 7.40 \\
\end{longtable}
\endgroup

%% file: sections/Table_Cubature_Stats_Longtable.tex
% Start alternating row colors on row 2 (header line is row 1)
\rowcolors{2}{white}{lightcornflower}

\begingroup
\setlength{\tabcolsep}{9pt} % PGK: eat up that white space!

\begin{longtable}{lrrrrrr}
    % --- 1. CAPTION & HEADER FOR FIRST PAGE ---
    \caption{
    \textbf{Cubature statistics.} 
    Error thresholds, tet counts, percentage, error, and elapsed times for our simulations.} 
    \label{tab:cubature_stats} \\
    \toprule
    \textbf{Example} & \textbf{Error Threshold} & \textbf{Num Modes} & \textbf{Cubature Tets} & \textbf{Percentage} & \textbf{Error} & \textbf{Elapsed Time (min)} \\
    \midrule
    \endfirsthead

    % --- 2. HEADER FOR SUBSEQUENT PAGES ---
    \caption*{(continued)} \\
    \toprule
    \textbf{Example} & \textbf{Error Threshold} & \textbf{Num Modes} & \textbf{Cubature Tets} & \textbf{Percentage} & \textbf{Error} & \textbf{Elapsed Time (min)} \\
    \midrule
    \endhead

    % --- 3. FOOTER FOR ALL PAGES EXCEPT LAST ---
    \midrule\rowcolor{white}
    \multicolumn{7}{r}{\small\itshape Continued on next page...} \\
    \endfoot

    % --- 4. FOOTER FOR LAST PAGE ---
    \bottomrule
    \endlastfoot

    % --- MERGED TABLE DATA ---
    armadillo & 0.05 & 25 & 3,564 & 5.23\% & 0.00142 & 2.43 \\
    armadillo & 0.05 & 60 & 13,350 & 19.61\% & 0.00394 & 5.98 \\
    armadillo & 0.025 & 25 & 7,460 & 10.96\% & 0.00514 & 4.07 \\
    armadillo & 0.025 & 60 & 23,538 & 34.57\% & 0.00279 & 8.33 \\
    beam & 0.05 & 25 & 7,342 & 22.38\% & 0.00228 & 3.02 \\
    beam & 0.05 & 60 & 19,137 & 58.34\% & 0.00408 & 5.13 \\
    beam & 0.025 & 25 & 14,029 & 42.76\% & 0.00286 & 5.66 \\
    beam & 0.025 & 60 & 22,221 & 67.74\% & 0.00472 & 5.29 \\
    bridge & 0.05 & 25 & 4,148 & 0.57\% & 0.02280 & 9.43 \\
    bridge & 0.05 & 60 & 8,257 & 1.14\% & 0.00130 & 7.50 \\
    bridge & 0.025 & 25 & 6,604 & 0.91\% & 0.00797 & 8.65 \\
    bridge & 0.025 & 60 & 12,366 & 1.70\% & 0.00080 & 3.71 \\
    chimpanzee & 0.05 & 25 & 1,130 & 5.62\% & 0.00209 & 0.20 \\
    chimpanzee & 0.05 & 60 & 3,322 & 16.52\% & 0.00242 & 0.31 \\
    chimpanzee & 0.025 & 25 & 2,082 & 10.35\% & 0.00406 & 0.26 \\
    chimpanzee & 0.025 & 60 & 5,112 & 25.42\% & 0.00468 & 0.39 \\
    coarse\_bunny & 0.05 & 25 & 1,393 & 61.26\% & 0.00642 & 0.13 \\
    coarse\_bunny & 0.05 & 60 & 1,696 & 74.58\% & 0.00732 & 0.14 \\
    coarse\_bunny & 0.025 & 25 & 1,664 & 73.18\% & 0.00653 & 0.14 \\
    coarse\_bunny & 0.025 & 60 & 1,780 & 78.28\% & 0.00711 & 0.14 \\
    dragon & 0.05 & 25 & 4,520 & 2.65\% & 0.00212 & 1.11 \\
    dragon & 0.05 & 60 & 18,803 & 11.03\% & 0.00224 & 2.10 \\
    dragon & 0.025 & 25 & 11,195 & 6.57\% & 0.00169 & 1.59 \\
    dragon & 0.025 & 60 & 37,829 & 22.18\% & 0.00244 & 3.18 \\
    ethernet\_cable & 0.05 & 25 & 253 & 0.48\% & 0.00146 & 0.20 \\
    ethernet\_cable & 0.05 & 60 & 7,329 & 13.96\% & 0.00150 & 0.68 \\
    ethernet\_cable & 0.025 & 25 & 1,557 & 2.96\% & 0.01050 & 0.37 \\
    ethernet\_cable & 0.025 & 60 & 17,805 & 33.91\% & 0.00239 & 1.29 \\
    gecko & 0.05 & 25 & 1,510 & 10.60\% & 0.00218 & 0.22 \\
    gecko & 0.05 & 60 & 3,952 & 27.73\% & 0.00247 & 0.35 \\
    gecko & 0.025 & 25 & 2,594 & 18.20\% & 0.00221 & 0.29 \\
    gecko & 0.025 & 60 & 6,060 & 42.53\% & 0.00266 & 0.57 \\
    meta\_basketball & 0.05 & 25 & 24,633 & 3.09\% & 0.00129 & 5.64 \\
    meta\_basketball & 0.05 & 60 & 46,420 & 5.83\% & 0.00125 & 7.61 \\
    meta\_basketball & 0.025 & 25 & 49,288 & 6.19\% & 0.00119 & 5.71 \\
    meta\_basketball & 0.025 & 60 & 85,201 & 10.70\% & 0.00142 & 7.45 \\
    rose & 0.05 & 25 & 171 & 0.10\% & 0.00081 & 1.00 \\
    rose & 0.05 & 60 & 469 & 0.26\% & 0.00136 & 1.62 \\
    rose & 0.025 & 25 & 271 & 0.15\% & 0.00052 & 1.25 \\
    rose & 0.025 & 60 & 486 & 0.27\% & 0.00053 & 2.12 \\
    rubber\_chicken & 0.05 & 25 & 3,209 & 7.52\% & 0.00558 & 0.52 \\
    rubber\_chicken & 0.05 & 60 & 12,164 & 28.50\% & 0.00246 & 1.11 \\
    rubber\_chicken & 0.025 & 25 & 6,348 & 14.87\% & 0.00180 & 0.70 \\
    rubber\_chicken & 0.025 & 60 & 19,729 & 46.22\% & 0.00366 & 1.61 \\
    schwartz & 0.05 & 25 & 24,125 & 3.69\% & 0.00087 & 4.06 \\
    schwartz & 0.05 & 60 & 154,000 & 23.53\% & 0.00184 & 11.35 \\
    schwartz & 0.025 & 25 & 140,088 & 21.41\% & 0.00183 & 10.79 \\
    schwartz & 0.025 & 60 & 199,162 & 30.44\% & 0.00233 & 13.60 \\
    shiba & 0.05 & 25 & 5,804 & 15.45\% & 0.00212 & 0.57 \\
    shiba & 0.05 & 60 & 15,308 & 40.74\% & 0.00341 & 1.17 \\
    shiba & 0.025 & 25 & 10,999 & 29.28\% & 0.00354 & 0.90 \\
    shiba & 0.025 & 60 & 20,384 & 54.25\% & 0.00321 & 1.35 \\
    strawberry & 0.05 & 25 & 10,033 & 3.63\% & 0.00544 & 1.99 \\
    strawberry & 0.05 & 60 & 27,979 & 10.13\% & 0.00359 & 3.33 \\
    strawberry & 0.025 & 25 & 15,147 & 5.48\% & 0.00426 & 2.52 \\
    strawberry & 0.025 & 60 & 52,041 & 18.84\% & 0.00313 & 4.76 \\
    tumbleweed & 0.05 & 25 & 8,139 & 3.69\% & 0.00077 & 1.94 \\
    tumbleweed & 0.05 & 60 & 9,773 & 4.43\% & 0.00092 & 1.87 \\
    tumbleweed & 0.025 & 25 & 12,630 & 5.72\% & 0.00096 & 1.90 \\
    tumbleweed & 0.025 & 60 & 15,454 & 7.00\% & 0.00090 & 2.09 \\
\end{longtable}
\endgroup

%% file: sections/Table_Perf_LongTable.tex
\begingroup
  % 10 columns need slightly tighter padding (5pt) to avoid overfull \hbox
  \setlength{\tabcolsep}{5pt}
  \rowcolors{2}{white}{lightcornflower}

  \begin{longtable}{lrrrrrrrrr}
    % --- 1. CAPTION & HEADER FOR FIRST PAGE ---
    \caption{
    \textbf{Performance.}
    Runtime in \textbf{milliseconds} with material parameters, error thresholds, and mode counts. Note that for all but our most complicated examples, every simulation runs at 100 steps/s. The $\Delta t$ time step for all simulations was $0.01$. Conjugate gradient absolute tolerance was $1e^{-6}$, run to convergence for each timestep.} 
    \label{tab:performance_stats_updated} \\
    \toprule
    \textbf{Example} & \textbf{Material} & \textbf{YM (Pa)} & \textbf{Error} & \textbf{Num Modes} & \textbf{N Steps} & \textbf{Total (ms)} & \textbf{Assembly (ms)} & \textbf{Floor (ms)} & \textbf{Solve (ms)} \\
    \midrule
    \endfirsthead

    % --- 2. HEADER FOR SUBSEQUENT PAGES ---
    \caption*{(continued)} \\
    \toprule
    \textbf{Example} & \textbf{Material} & \textbf{YM (Pa)} & \textbf{Error} & \textbf{Num Modes} & \textbf{N Steps} & \textbf{Total (ms)} & \textbf{Assembly (ms)} & \textbf{Floor (ms)} & \textbf{Solve (ms)} \\
    \midrule
    \endhead

    % --- 3. FOOTER FOR ALL PAGES EXCEPT LAST ---
    \midrule\rowcolor{white}
    \multicolumn{10}{r}{\small\itshape Continued on next page...} \\
    \endfoot

    % --- 4. FOOTER FOR LAST PAGE ---
    \bottomrule
    \endlastfoot

    % --- TABLE DATA ---
    armadillo & ARAP & $10^{7}$ & 0.05 & 25 & 419 & 2.86 $\pm$ 0.57 & 0.11 $\pm$ 0.33 & 0.19 $\pm$ 0.16 & 1.62 $\pm$ 0.09 \\
    basketball & ARAP & $10^{7}$ & 0.05 & 25 & 1,499 & 6.64 $\pm$ 1.96 & 0.13 $\pm$ 0.27 & 1.63 $\pm$ 1.80 & 1.33 $\pm$ 0.15 \\
    beam & ARAP & $10^{7}$ & 0.05 & 25 & 599 & 3.66 $\pm$ 0.82 & 0.12 $\pm$ 0.34 & 1.13 $\pm$ 0.53 & 1.01 $\pm$ 0.11 \\
    bridge & ARAP & $10^{7}$ & 0.025 & 60 & 1,999 & 13.20 $\pm$ 1.72 & 0.22 $\pm$ 0.45 & 0.84 $\pm$ 0.16 & 10.25 $\pm$ 1.47 \\
    chimpanzee & ARAP & $10^{7}$ & 0.05 & 25 & 999 & 2.19 $\pm$ 0.30 & 0.07 $\pm$ 0.17 & 0.20 $\pm$ 0.05 & 1.47 $\pm$ 0.18 \\
    coarse\_bunny & ARAP & $10^{7}$ & 0.05 & 25 & 999 & 5.33 $\pm$ 1.43 & 0.06 $\pm$ 0.10 & 0.22 $\pm$ 0.11 & 4.54 $\pm$ 1.40 \\
    dragon & ARAP & $10^{7}$ & 0.05 & 25 & 1,999 & 6.49 $\pm$ 3.38 & 0.07 $\pm$ 0.19 & 4.17 $\pm$ 3.34 & 1.26 $\pm$ 0.10 \\
    ethernet\_cable & ARAP & $10^{7}$ & 0.05 & 25 & 999 & 1.83 $\pm$ 0.47 & 0.06 $\pm$ 0.13 & 0.23 $\pm$ 0.08 & 1.14 $\pm$ 0.23 \\
    gecko & ARAP & $10^{7}$ & 0.05 & 25 & 2,999 & 2.22 $\pm$ 0.40 & 0.07 $\pm$ 0.16 & 0.49 $\pm$ 0.21 & 1.09 $\pm$ 0.08 \\
    rose & ARAP & $10^{8}$ & 0.05 & 25 & 999 & 1.38 $\pm$ 0.43 & 0.05 $\pm$ 0.07 & 0.25 $\pm$ 0.13 & 0.74 $\pm$ 0.16 \\
    rubber\_chicken & ARAP & $10^{6}$ & 0.05 & 25 & 1,499 & 2.30 $\pm$ 0.58 & 0.11 $\pm$ 0.33 & 0.23 $\pm$ 0.16 & 1.23 $\pm$ 0.25 \\
    rubber\_chicken & ARAP & $10^{8}$ & 0.05 & 25 & 299 & 3.29 $\pm$ 0.43 & 0.08 $\pm$ 0.19 & 0.19 $\pm$ 0.08 & 2.16 $\pm$ 0.19 \\
    schwartz & ARAP & $10^{7}$ & 0.025 & 25 & 999 & 7.19 $\pm$ 0.84 & 0.07 $\pm$ 0.15 & 0.92 $\pm$ 0.20 & 1.21 $\pm$ 0.23 \\
    shiba & ARAP & $10^{7}$ & 0.05 & 25 & 999 & 2.55 $\pm$ 0.88 & 0.13 $\pm$ 0.36 & 0.23 $\pm$ 0.12 & 0.88 $\pm$ 0.71 \\
    shiba & ARAP & $10^{7}$ & 0.05 & 60 & 999 & 5.36 $\pm$ 1.46 & 0.20 $\pm$ 0.31 & 0.37 $\pm$ 0.11 & 1.24 $\pm$ 1.26 \\
    strawberry & ARAP & $10^{6}$ & 0.05 & 35 & 999 & 4.16 $\pm$ 0.69 & 0.07 $\pm$ 0.16 & 0.50 $\pm$ 0.16 & 1.02 $\pm$ 0.15 \\
    tumbleweed & ARAP & $10^{7}$ & 0.025 & 25 & 999 & 10.75 $\pm$ 1.43 & 0.24 $\pm$ 0.50 & 2.05 $\pm$ 0.58 & 6.26 $\pm$ 1.03 \\
    tumbleweed & ARAP & $10^{7}$ & 0.025 & 60 & 999 & 18.15 $\pm$ 2.01 & 0.20 $\pm$ 0.30 & 1.68 $\pm$ 0.34 & 13.95 $\pm$ 1.85 \\
    tumbleweed & ARAP & $10^{7}$ & 0.05 & 25 & 999 & 6.40 $\pm$ 1.26 & 0.24 $\pm$ 0.53 & 2.01 $\pm$ 0.66 & 2.18 $\pm$ 0.24 \\
    tumbleweed & ARAP & $10^{7}$ & 0.05 & 60 & 999 & 6.49 $\pm$ 1.31 & 0.28 $\pm$ 0.56 & 2.11 $\pm$ 0.56 & 2.03 $\pm$ 0.19 \\
  \end{longtable}
\endgroup

%% file: sections/Appendix_Screengrabs.tex
\begin{figure*}[ht]
\includegraphics[width=\textwidth]{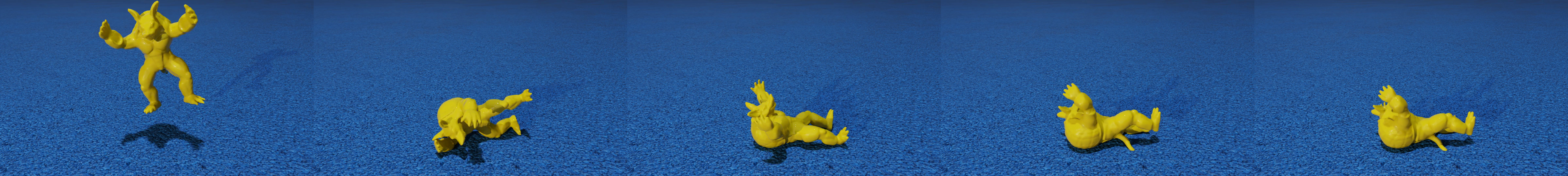}
\Description{Images from video.}\caption{Screengrabs from rendered, real-time simulations using decimated handles.}
\label{fig:armadillo}
 \end{figure*}
 \begin{figure*}[ht]
\includegraphics[width=\textwidth]{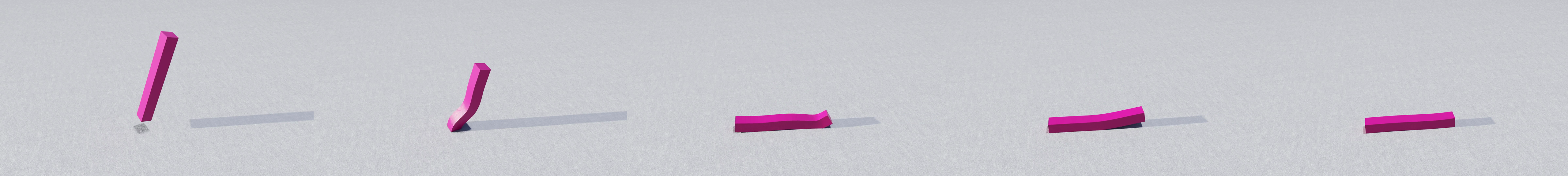}
\Description{Images from video.}\caption{Screengrabs from rendered, real-time simulations using decimated handles.}
\label{fig:beam}
 \end{figure*}
 \begin{figure*}[ht]
\includegraphics[width=\textwidth]{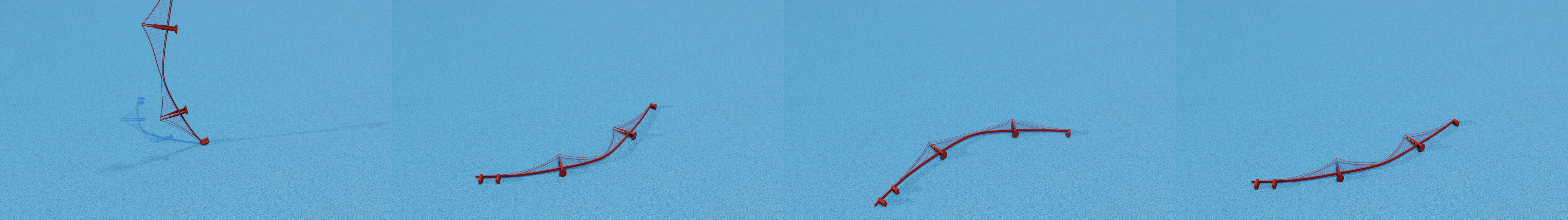}
\Description{Images from video.}\caption{Screengrabs from rendered, real-time simulations using decimated handles.}
\label{fig:bridge}
 \end{figure*}
 \begin{figure*}[ht]
\includegraphics[width=\textwidth]{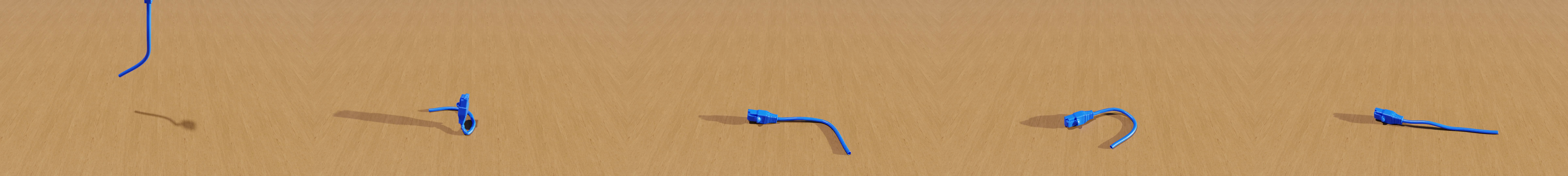}
\Description{Images from video.}\caption{Screengrabs from rendered, real-time simulations using decimated handles.}
\label{fig:cable}
 \end{figure*}
 \begin{figure*}[ht]
\includegraphics[width=\textwidth]{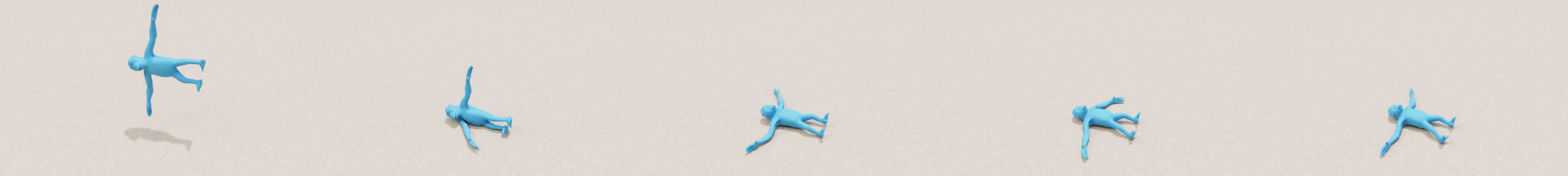}
\Description{Images from video.}\caption{Screengrabs from rendered, real-time simulations using decimated handles.}
\label{fig:chimp}
 \end{figure*}
 \begin{figure*}[ht]
\includegraphics[width=\textwidth]{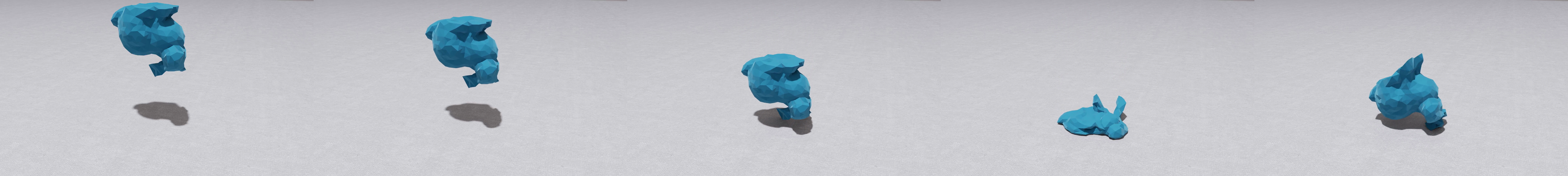}
\Description{Images from video.}\caption{Screengrabs from rendered, real-time simulations using decimated handles.}
\label{fig:bunny}
 \end{figure*}
 \begin{figure*}[ht]
\includegraphics[width=\textwidth]{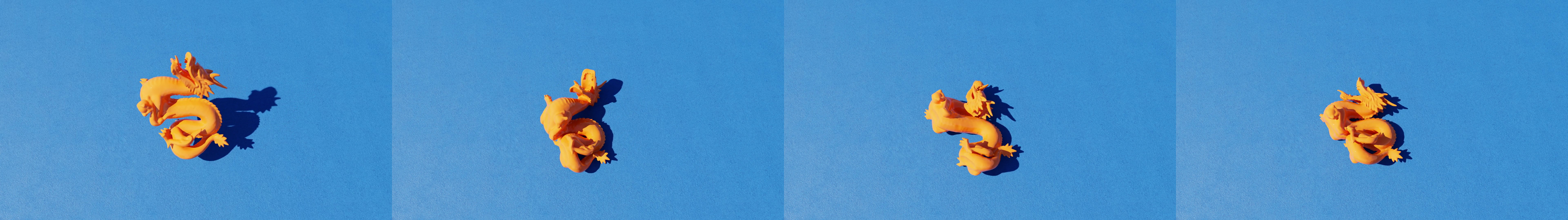}
\Description{Images from video.}\caption{Screengrabs from rendered, real-time simulations using decimated handles.}
\label{fig:dragon}
\end{figure*}

\begin{figure*}[ht]
\includegraphics[width=\textwidth]{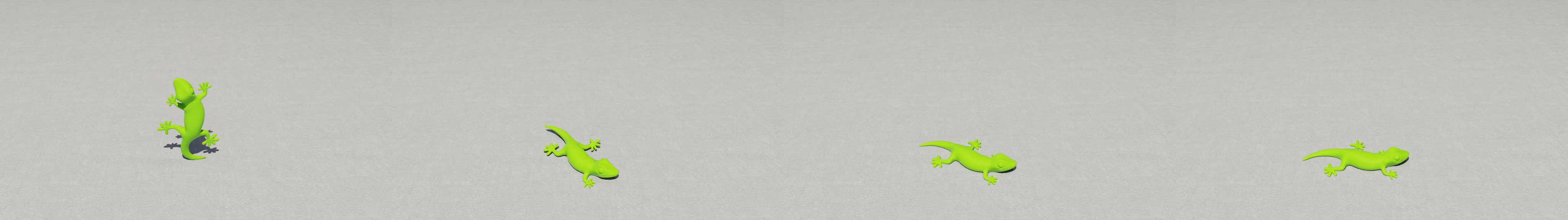}
\Description{Images from video.}\caption{Screengrabs from rendered, real-time simulations using decimated handles.}
\label{fig:gecko}
\end{figure*}

\begin{figure*}[ht]
\includegraphics[width=\textwidth]{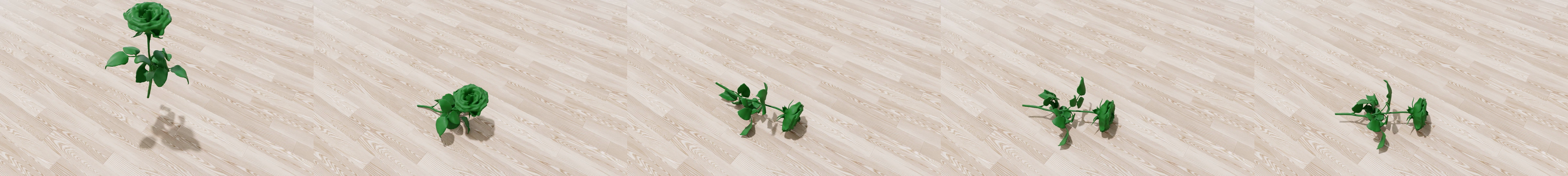}
\Description{Images from video.}\caption{Screengrabs from rendered, real-time simulations using decimated handles.}
\label{fig:rose}
\end{figure*}

\begin{figure*}[ht]
\includegraphics[width=\textwidth]{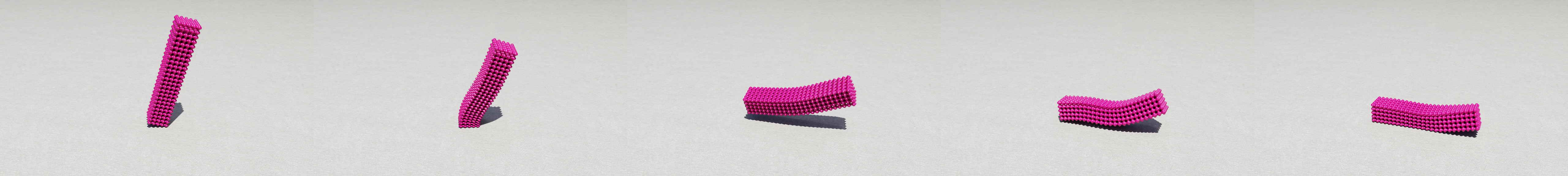}
\Description{Images from video.}\caption{Screengrabs from rendered, real-time simulations using decimated handles.}
\label{fig:schwartz}
\end{figure*}

\begin{figure*}[ht]
\includegraphics[width=\textwidth]{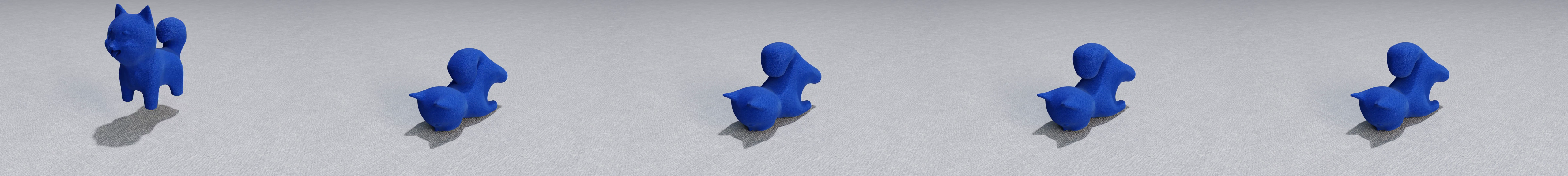}
\Description{Images from video.}\caption{Screengrabs from rendered, real-time simulations using decimated handles.}
\label{fig:shiba}
\end{figure*}

\begin{figure*}[ht]
\includegraphics[width=\textwidth]{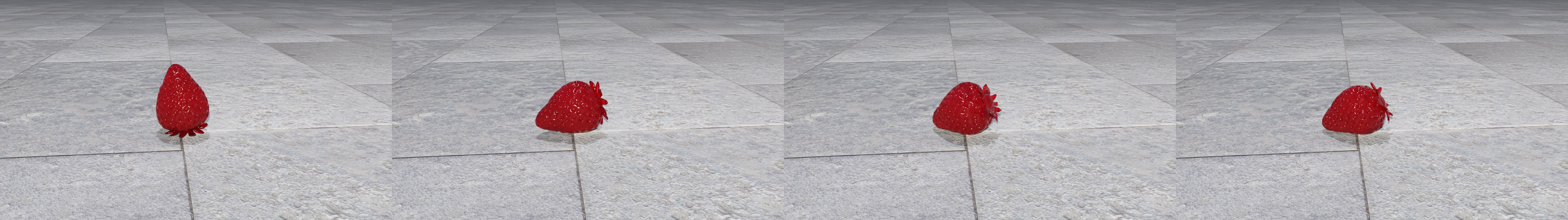}
\Description{Images from video.}\caption{Screengrabs from rendered, real-time simulations using decimated handles.}
\label{fig:strawberry}
\end{figure*}

\begin{figure*}[ht]
\includegraphics[width=\textwidth]{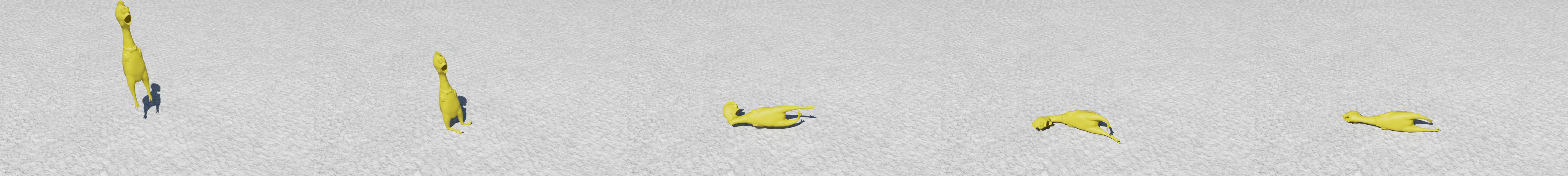}
\Description{Images from video.}\caption{Rubber chicken simulation with stiffness of $10^{6}$ Pa.}
\label{fig:chicken1}
\end{figure*}

\begin{figure*}[ht]
\includegraphics[width=\textwidth]{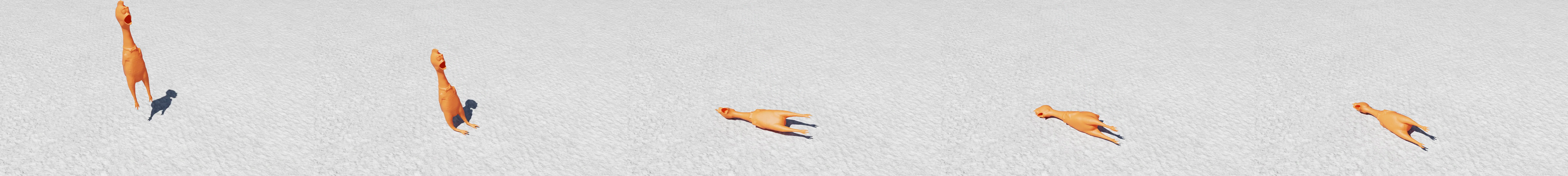}
\Description{Images from video.}\caption{Rubber chicken simulation with stiffness of $10^{8}$ Pa.}
\label{fig:chicken2}
\end{figure*}

%% file: lovehandles.bib
@article{PentlandWilliams1989,
author = {Pentland, A. and Williams, J.},
title = {Good vibrations: modal dynamics for graphics and animation},
year = {1989},
issue_date = {July 1989},
publisher = {Association for Computing Machinery},
address = {New York, NY, USA},
volume = {23},
number = {3},
issn = {0097-8930},
url = {https://doi.org/10.1145/74334.74355},
doi = {10.1145/74334.74355},
journal = {SIGGRAPH Comput. Graph.},
month = jul,
pages = {207–214},
numpages = {8}
}

@article{James2005,
author = {James, Doug L. and Twigg, Christopher D.},
title = {Skinning mesh animations},
year = {2005},
issue_date = {July 2005},
publisher = {Association for Computing Machinery},
address = {New York, NY, USA},
volume = {24},
number = {3},
issn = {0730-0301},
url = {https://doi.org/10.1145/1073204.1073206},
doi = {10.1145/1073204.1073206},
journal = {ACM Trans. Graph.},
pages = {399-–407},
numpages = {9}
}

@article{an2008optimizing,
  title={Optimizing cubature for efficient integration of subspace deformations},
  author={An, Steven S and Kim, Theodore and James, Doug L},
  journal={ACM transactions on graphics (TOG)},
  volume={27},
  number={5},
  pages={1--10},
  year={2008},
  publisher={ACM New York, NY, USA}
}

@article{Gilles2011,
author = {Gilles, Benjamin and Bousquet, Guillaume and Faure, Francois and Pai, Dinesh K.},
title = {Frame-based elastic models},
year = {2011},
issue_date = {April 2011},
publisher = {Association for Computing Machinery},
address = {New York, NY, USA},
volume = {30},
number = {2},
issn = {0730-0301},
url = {https://doi.org/10.1145/1944846.1944855},
doi = {10.1145/1944846.1944855},
journal = {ACM Trans. Graph.},
month = apr,
articleno = {15},
numpages = {12}
}

@article{Faure2011,
author = {Faure, Fran\c{c}ois and Gilles, Benjamin and Bousquet, Guillaume and Pai, Dinesh K.},
title = {Sparse meshless models of complex deformable solids},
year = {2011},
issue_date = {July 2011},
publisher = {Association for Computing Machinery},
address = {New York, NY, USA},
volume = {30},
number = {4},
issn = {0730-0301},
url = {https://doi.org/10.1145/2010324.1964968},
doi = {10.1145/2010324.1964968},
journal = {ACM Trans. Graph.},
month = jul,
articleno = {73},
numpages = {10}
}

@inproceedings{Garland1997,
author = {Garland, Michael and Heckbert, Paul S.},
title = {Surface simplification using quadric error metrics},
year = {1997},
isbn = {0897918967},
publisher = {ACM Press/Addison-Wesley Publishing Co.},
address = {USA},
url = {https://doi.org/10.1145/258734.258849},
doi = {10.1145/258734.258849},
booktitle = {Proceedings of the 24th Annual Conference on Computer Graphics and Interactive Techniques},
pages = {209–216},
numpages = {8},
series = {SIGGRAPH '97}
}

@article{Simplicits2024,
author = {Modi, Vismay and Sharp, Nicholas and Perel, Or and Sueda, Shinjiro and Levin, David I. W.},
title = {Simplicits: Mesh-Free, Geometry-Agnostic Elastic Simulation},
year = {2024},
issue_date = {July 2024},
publisher = {Association for Computing Machinery},
address = {New York, NY, USA},
volume = {43},
number = {4},
issn = {0730-0301},
url = {https://doi.org/10.1145/3658184},
doi = {10.1145/3658184},
journal = {ACM Trans. Graph.},
month = jul,
articleno = {117},
numpages = {11}
}

@article{Barbic2005,
author = {Barbi\v{c}, Jernej and James, Doug L.},
title = {Real-Time subspace integration for St. Venant-Kirchhoff deformable models},
year = {2005},
issue_date = {July 2005},
publisher = {Association for Computing Machinery},
address = {New York, NY, USA},
volume = {24},
number = {3},
issn = {0730-0301},
url = {https://doi.org/10.1145/1073204.1073300},
doi = {10.1145/1073204.1073300},
journal = {ACM Trans. Graph.},
month = jul,
pages = {982–990},
numpages = {9}
}

@ARTICLE{ChoiKo2005,
  author={Min Gyu Choi and Hyeong-Seok Ko},
  journal={IEEE Transactions on Visualization and Computer Graphics}, 
  title={Modal warping: real-time simulation of large rotational deformation and manipulation}, 
  year={2005},
  volume={11},
  number={1},
  pages={91-101},
  doi={10.1109/TVCG.2005.13}}

@article{osqp,
  author  = {Stellato, B. and Banjac, G. and Goulart, P. and Bemporad, A. and Boyd, S.},
  title   = {{OSQP}: an operator splitting solver for quadratic programs},
  journal = {Mathematical Programming Computation},
  volume  = {12},
  number  = {4},
  pages   = {637--672},
  year    = {2020},
  doi     = {10.1007/s12532-020-00179-2},
  url     = {https://doi.org/10.1007/s12532-020-00179-2},
}

@article{yen1970algorithm,
  title={An algorithm for finding shortest routes from all source nodes to a given destination in general networks},
  author={Yen, Jin Y},
  journal={Quarterly of applied mathematics},
  volume={27},
  number={4},
  pages={526--530},
  year={1970}
}

@article{bellman1958routing,
  title={On a routing problem},
  author={Bellman, Richard},
  journal={Quarterly of applied mathematics},
  volume={16},
  number={1},
  pages={87--90},
  year={1958}
}

@article{benchekroun2023fast,
  title={Fast Complementary Dynamics via Skinning Eigenmodes},
  author={Benchekroun, Otman and Zhang, Jiayi Eris and Chaudhuri, Siddartha and Grinspun, Eitan and Zhou, Yi and Jacobson, Alec},
  journal={ACM Transactions on Graphics (TOG)},
  volume={42},
  number={4},
  pages={1--21},
  year={2023},
  publisher={ACM New York, NY, USA}
}

@article{brandt2017compressed,
  title={Compressed vibration modes of elastic bodies},
  author={Brandt, Christopher and Hildebrandt, Klaus},
  journal={Computer Aided Geometric Design},
  volume={52},
  pages={297--312},
  year={2017},
  publisher={Elsevier}
}

@article{sellan2023breaking,
  title={Breaking good: Fracture modes for realtime destruction},
  author={Sell{\'a}n, Silvia and Luong, Jack and Mattos Da Silva, Leticia and Ramakrishnan, Aravind and Yang, Yuchuan and Jacobson, Alec},
  journal={ACM Transactions on Graphics},
  volume={42},
  number={1},
  pages={1--12},
  year={2023},
  publisher={ACM New York, NY}
}

@article{benchekroun2025force,
  title={Force-Dual Modes: Subspace Design from Stochastic Forces},
  author={Benchekroun, Otman and Grinspun, Eitan and Chiaramonte, Maurizio and Etter, Philip Allen},
  journal={ACM Transactions on Graphics (TOG)},
  volume={44},
  number={6},
  pages={1--14},
  year={2025},
  publisher={ACM New York, NY, USA}
}

@inproceedings{chang2023licrom,
author = {Chang, Yue and Chen, Peter Yichen and Wang, Zhecheng and Chiaramonte, Maurizio M. and Carlberg, Kevin and Grinspun, Eitan},
title = {Li{CROM}: Linear-Subspace Continuous Reduced Order Modeling with Neural Fields},
year = {2023},
isbn = {9798400703157},
publisher = {Association for Computing Machinery},
address = {New York, NY, USA},
url = {https://doi.org/10.1145/3610548.3618158},
doi = {10.1145/3610548.3618158},
booktitle = {SIGGRAPH Asia 2023 Conference Papers},
articleno = {111},
numpages = {12},
location = {Sydney, NSW, Australia},
series = {SA '23}
}

@inproceedings{xiang2026freeform,
  title={FreeForm: Reduced-Order Deformable Simulation from Particle-Based Skinning Eigenmodes},
  author={Xiang, Donglai and Modi, Vismay and Dagli, Rishit and Trusty, Ty and Daviet, Gilles and Chen, Anka He and Sharp, Nicholas and Levin, David IW},
  booktitle={Proceedings of the IEEE/CVF Conference on Computer Vision and Pattern Recognition},
  pages={32475--32484},
  year={2026}
}

@article{trusty2025sparse,
  title={Sparse, Geometry-and Material-Aware Bases for Multilevel Elastodynamic Simulation},
  author={Trusty, Ty and Levin, David IW and Kaufman, Danny M},
  journal={arXiv preprint arXiv:2508.13386},
  year={2025}
}

@article{chen2019material,
  title={Material-adapted refinable basis functions for elasticity simulation},
  author={Chen, Jiong and Budninskiy, Max and Owhadi, Houman and Bao, Hujun and Huang, Jin and Desbrun, Mathieu},
  journal={ACM Transactions on Graphics (TOG)},
  volume={38},
  number={6},
  pages={1--15},
  year={2019},
  publisher={ACM New York, NY, USA}
}

@article{chen2017dynamics,
author = {Chen, Desai and Levin, David I. W. and Matusik, Wojciech and Kaufman, Danny M.},
title = {Dynamics-aware numerical coarsening for fabrication design},
year = {2017},
issue_date = {August 2017},
publisher = {Association for Computing Machinery},
address = {New York, NY, USA},
volume = {36},
number = {4},
issn = {0730-0301},
url = {https://doi.org/10.1145/3072959.3073669},
doi = {10.1145/3072959.3073669},
journal = {ACM Trans. Graph.},
month = jul,
articleno = {84},
numpages = {15}
}

@article{xian2019scalable,
  title={A scalable galerkin multigrid method for real-time simulation of deformable objects},
  author={Xian, Zangyueyang and Tong, Xin and Liu, Tiantian},
  journal={ACM Transactions on Graphics (TOG)},
  volume={38},
  number={6},
  pages={1--13},
  year={2019},
  publisher={ACM New York, NY, USA}
}

@inproceedings{trusty2023subspace,
author = {Trusty, Ty and Benchekroun, Otman and Grinspun, Eitan and Kaufman, Danny M. and Levin, David I.W.},
title = {Subspace Mixed Finite Elements for Real-Time Heterogeneous Elastodynamics},
year = {2023},
isbn = {9798400703157},
publisher = {Association for Computing Machinery},
address = {New York, NY, USA},
url = {https://doi.org/10.1145/3610548.3618220},
doi = {10.1145/3610548.3618220},
booktitle = {SIGGRAPH Asia 2023 Conference Papers},
articleno = {112},
numpages = {10},
location = {Sydney, NSW, Australia},
series = {SA '23}
}

@article{kheradmand20243d,
  title={3d gaussian splatting as markov chain monte carlo},
  author={Kheradmand, Shakiba and Rebain, Daniel and Sharma, Gopal and Sun, Weiwei and Tseng, Yang-Che and Isack, Hossam and Kar, Abhishek and Tagliasacchi, Andrea and Yi, Kwang Moo},
  journal={Advances in Neural Information Processing Systems},
  volume={37},
  pages={80965--80986},
  year={2024}
}

@inproceedings{baraff2023large,
author = {Baraff, David and Witkin, Andrew},
title = {Large steps in cloth simulation},
year = {1998},
isbn = {0897919998},
publisher = {Association for Computing Machinery},
address = {New York, NY, USA},
url = {https://doi.org/10.1145/280814.280821},
doi = {10.1145/280814.280821},
booktitle = {Proceedings of the 25th Annual Conference on Computer Graphics and Interactive Techniques},
pages = {43–54},
numpages = {12},
series = {SIGGRAPH '98}
}

@misc{warp2022,
  title        = {Warp: A High-performance Python Framework for GPU Simulation and Graphics},
  author       = {Miles Macklin},
  month        = {March},
  year         = {2022},
  note         = {NVIDIA GPU Technology Conference (GTC)},
  howpublished = {\url{https://github.com/nvidia/warp}}
}

@ARTICLE{2020SciPy-NMeth,
  author  = {Virtanen, Pauli and Gommers, Ralf and Oliphant, Travis E. and
            Haberland, Matt and Reddy, Tyler and Cournapeau, David and
            Burovski, Evgeni and Peterson, Pearu and Weckesser, Warren and
            Bright, Jonathan and {van der Walt}, St{\'e}fan J. and
            Brett, Matthew and Wilson, Joshua and Millman, K. Jarrod and
            Mayorov, Nikolay and Nelson, Andrew R. J. and Jones, Eric and
            Kern, Robert and Larson, Eric and Carey, C J and
            Polat, {\.I}lhan and Feng, Yu and Moore, Eric W. and
            {VanderPlas}, Jake and Laxalde, Denis and Perktold, Josef and
            Cimrman, Robert and Henriksen, Ian and Quintero, E. A. and
            Harris, Charles R. and Archibald, Anne M. and
            Ribeiro, Ant{\^o}nio H. and Pedregosa, Fabian and
            {van Mulbregt}, Paul and {SciPy 1.0 Contributors}},
  title   = {{{SciPy} 1.0: Fundamental Algorithms for Scientific
            Computing in Python}},
  journal = {Nature Methods},
  year    = {2020},
  volume  = {17},
  pages   = {261--272},
  adsurl  = {https://rdcu.be/b08Wh},
  doi     = {10.1038/s41592-019-0686-2},
}

@misc{libigl,
  title = { {libigl}: A simple {C++} geometry processing library},
  author = {Alec Jacobson and Daniele Panozzo and others},
  note = {https://libigl.github.io/},
  year = {2018},
}

@misc{ovrtx_sdk,
  title        = {{{ovrtx}: A C and Python library for physically accurate, real-time, sensor simulation and visualization using NVIDIA Omniverse RTX}},
  author       = {{NVIDIA Omniverse Development Team}},
  year         = {2026},
  howpublished = {\url{https://github.com/nvidia-omniverse/ovrtx}},
  note         = {NVIDIA Corporation}
}

@misc{omniverse_kit_app_template,
  title        = {{{NVIDIA Omniverse Kit App Template}}},
  author       = {{NVIDIA Corporation}},
  year         = {2026},
  howpublished = {\url{https://github.com/NVIDIA-Omniverse/kit-app-template}},
  note         = {Available as a GitHub repository for OpenUSD-based application development using Omniverse Kit SDK}
}

@misc{dong2024rayleighquotientgraphneural,
      title={Rayleigh Quotient Graph Neural Networks for Graph-level Anomaly Detection}, 
      author={Xiangyu Dong and Xingyi Zhang and Sibo Wang},
      year={2024},
      eprint={2310.02861},
      archivePrefix={arXiv},
      primaryClass={cs.LG},
      url={https://arxiv.org/abs/2310.02861}, 
}

@article{10.1145/3197517.3201387, author = {Brandt, Christopher and Eisemann, Elmar and Hildebrandt, Klaus}, title = {Hyper-reduced projective dynamics}, year = {2018}, issue_date = {August 2018}, publisher = {Association for Computing Machinery}, address = {New York, NY, USA}, volume = {37}, number = {4}, issn = {0730-0301}, url = {https://doi.org/10.1145/3197517.3201387}, doi = {10.1145/3197517.3201387}, journal = {ACM Trans. Graph.}, month = jul, articleno = {80}, numpages = {13} }

@inproceedings{liao2026boundaryaware,
author = {Liao, Li and Shen, Pengfei and Peng, Yifan},
title = {Boundary-aware Neural Model Reduction for PDEs},
year = {2026},
isbn = {9798400725548},
publisher = {Association for Computing Machinery},
address = {New York, NY, USA},
url = {https://doi.org/10.1145/3799902.3811153},
doi = {10.1145/3799902.3811153},
booktitle = {Proceedings of the Special Interest Group on Computer Graphics and Interactive Techniques Conference Conference Papers},
articleno = {5},
numpages = {11},
location = {
},
series = {SIGGRAPH Conference Papers '26}
}

@inproceedings{
chen2023crom,
title={{CROM}: Continuous Reduced-Order Modeling of {PDE}s Using Implicit Neural Representations},
author={Peter Yichen Chen and Jinxu Xiang and Dong Heon Cho and Yue Chang and G A Pershing and Henrique Teles Maia and Maurizio M Chiaramonte and Kevin Thomas Carlberg and Eitan Grinspun},
booktitle={The Eleventh International Conference on Learning Representations },
year={2023},
url={https://openreview.net/forum?id=FUORz1tG8Og}
}

@article{10.1145/3450626.3459753,
author = {Lan, Lei and Yang, Yin and Kaufman, Danny and Yao, Junfeng and Li, Minchen and Jiang, Chenfanfu},
title = {Medial IPC: accelerated incremental potential contact with medial elastics},
year = {2021},
issue_date = {August 2021},
publisher = {Association for Computing Machinery},
address = {New York, NY, USA},
volume = {40},
number = {4},
issn = {0730-0301},
url = {https://doi.org/10.1145/3450626.3459753},
doi = {10.1145/3450626.3459753},
journal = {ACM Trans. Graph.},
month = jul,
articleno = {158},
numpages = {16}
}

@article{li2019multi,
  title={Multi-resolution modeling of shapes in contact},
  author={Li, Yijing and Barbi{\v{c}}, Jernej},
  journal={Proceedings of the ACM on Computer Graphics and Interactive Techniques},
  volume={2},
  number={2},
  pages={1--26},
  year={2019},
  publisher={ACM New York, NY, USA}
}

@article{Hu:2018:TMW:3197517.3201353,
 author = {Hu, Yixin and Zhou, Qingnan and Gao, Xifeng and Jacobson, Alec and Zorin, Denis and Panozzo, Daniele},
 title = {Tetrahedral Meshing in the Wild},
 journal = {ACM Trans. Graph.},
 issue_date = {August 2018},
 volume = {37},
 number = {4},
 month = jul,
 year = {2018},
 issn = {0730-0301},
 pages = {60:1--60:14},
 articleno = {60},
 numpages = {14},
 url = {http://doi.acm.org/10.1145/3197517.3201353},
 doi = {10.1145/3197517.3201353},
 acmid = {3201353},
 publisher = {ACM},
 address = {New York, NY, USA},
}

@article{vonTycowicz2013,
author = {von Tycowicz, Christoph and Schulz, Christian and Seidel, Hans-Peter and Hildebrandt, Klaus},
title = {An efficient construction of reduced deformable objects},
year = {2013},
issue_date = {November 2013},
publisher = {Association for Computing Machinery},
address = {New York, NY, USA},
volume = {32},
number = {6},
issn = {0730-0301},
url = {https://doi.org/10.1145/2508363.2508392},
doi = {10.1145/2508363.2508392},
journal = {ACM Trans. Graph.},
month = nov,
articleno = {213},
numpages = {10}
}

@inproceedings{Badler1982Modelling,
  author    = {Badler, N. I. and Morris, M.},
  title     = {Modelling flexible articulated objects},
  booktitle = {Proc. Computer Graphics' 82},
  year      = {1982},
  publisher = {Online Conferences},
  pages     = {305--314}
}
